\documentclass[aip,preprint,jap]{revtex4-1}

\usepackage[english]{babel}
\usepackage[T1]{fontenc}
\usepackage{graphicx}
\usepackage{pstricks}
\usepackage{pst-node}
\usepackage{amsmath}
\usepackage[normalem]{ulem}

\usepackage{graphicx}  
\usepackage{amsmath}   
\usepackage{url}
\usepackage{multirow}
\usepackage{pstricks}
\usepackage{pst-node}
\usepackage{multirow}
\usepackage{rotating}
\usepackage{amsfonts}
\usepackage{amssymb}
\usepackage{exscale}
\usepackage{subfigure}
\usepackage{upgreek}
\usepackage{textcomp}
\usepackage{epsfig}
\usepackage[T1]{fontenc}
\usepackage{perpage}

\newcommand{\rmd}{{\rm d}}

\begin{document}

%\preprint{}

\title{Ripple field losses in DC biased superconductors: simulations and comparison with measurements} %Title of paper



\author{Valtteri Lahtinen}
\email[]{valtteri.lahtinen@tut.fi}
%\homepage[]{}
%\thanks{}
\affiliation{Tampere University of Technology, Department of Electrical Engineering, Electromagnetics, PO Box 692, 33101 Tampere, Finland, http://www.tut.fi/smg }

\author{Enric Pardo}
\affiliation{Institute of Electrical Engineering, Slovak Academy of Sciences, Dubravska 9, 84104 Bratislava, Slovakia}

\author{Ján \v{S}ouc}
\affiliation{Institute of Electrical Engineering, Slovak Academy of Sciences, Dubravska 9, 84104 Bratislava, Slovakia}

\author{Mykola Solovyov}
\affiliation{Institute of Electrical Engineering, Slovak Academy of Sciences, Dubravska 9, 84104 Bratislava, Slovakia}

\author{Antti Stenvall}
\affiliation{Tampere University of Technology, Department of Electrical Engineering, Electromagnetics, PO Box 692, 33101 Tampere, Finland, http://www.tut.fi/smg }


\date{\today}

\begin{abstract}
For example in some supercondcuting generators, motors and power transmission cables the superconductor experiences a changing magnetic field in a DC background. Simulating the losses caused by this AC ripple field is an important task from the application design point of view. In this work, we compare two formulations, the $H$-formulation and the minimum magnetic energy variation (the MMEV-formulation), based on the eddy current model (ECM) and the critical state model (CSM), respectively, in simulating ripple field losses in a DC biased coated conductor tape. Furthermore, we compare our simulation results with measurements. We investigate the frequency-dependence of the hysteresis loss predictions of the power law based ECM and verify by a measurement, that in DC use, ECM clearly over-estimates the homogenization of the current density profile in the coated conductor tape: the relaxation of the local current density is not nearly as prominent in the measurement as it is in the simulation. Hence, we suggest that the power law resistivity, used as the \emph{local} relation between the electric field intensity ${\bf E}$ and current density ${\bf J}$ in ECM, is not an intrinsic property of high-temperature superconductors. The difference between the models manifests itself as discrepancies in ripple field loss simulations in very low AC fields with significant DC fields or currents involved. The results also show, however, that for many practical situations, CSM and ECM are both eligible models for ripple field loss simulations.
\end{abstract}



\maketitle %\maketitle must follow title, authors, abstract and \pacs

\section{Introduction}

In superconducting applications, such as motors, generators and some power transmission cables, superconductors experience changing magnetic fields while carrying DC currents -- sometimes also in a DC background field. From the simulation point of view, these DC-AC situations are inherently different from pure AC situations due to longer transients, and computing AC losses in superconductors in such cases should be performed with particular care. As AC losses can be a restricting factor for the feasibility and functionality of an application, simulating these so called ripple field losses is extremely important in the design phase to avoid unnecessary costs of constructing defunct prototypes not matching the specifications.

The two most widely used models for simulating hysteresis losses in superconductors are the critical state model (CSM)~\cite{bean62PRL,prigozhin97IES} and the eddy current model (ECM)~\cite{lahtinen13IES}. Both models employ the magnetoquasistatic approximation of the Maxwell's theory. The separating factor between the two models is the relation between the electric field intensity ${\bf E}$ and the current density ${\bf J}$. In CSM, the relation is assumed to be sharp, whereas in ECM, a smooth power law dependence, suggested by macroscopic measurement data~\cite{bruzzone04PhC}, is assumed between ${\bf E}$ and ${\bf J}$.

CSM and ECM, formulated in various ways, have both been shown to be eligible models for AC loss simulations in a wide range of situations (see \cite{norris70JPD,tapefmBi,HacIacinphase,pardo13IES,roebelcomp} and \cite{roebelcomp,stenvall10SSTb,nguyen10SST,hongZ11IES} for CSM and ECM results, respectively). However, the difference in their predictions for superconductors in low frequency use has not gained much attention. Especially in ripple field loss simulations in significant DC fields, the difference can be prominent \cite{lahtinen13IES}. In this paper, we show how the models differ when simulating a $Re$BCO coated conductor tape ($Re$BCO stands for $Re$Ba$_2$Cu$_3$O$_{7-x}$, where $Re$ is a rare-earth, typically Y, Gd or Sm.) in a simple DC transport current case and compare these results with measurements. Through simulations, also the frequency-dependence of hysteresis loss predictions of ECM is investigated. Furthermore, we compare CSM and ECM in ripple field loss simulations for a DC biased coated conductor tape and investigate the appropriateness of the models by comparing simulation results with AC loss measurements and magnetic field mapping.

In section~2 we introduce the models we are comparing and the formulations we use to present them. The methods of measurement we have used are discussed in section~3, and in section~4 the simulation and measurement results are presented and discussed. Finally, in section~5, conclusions are drawn.

\section{The models and the formulations}

The same theory can have many different models resulting in different descriptions of reality. AC loss simulations of superconductors are usually based on models of the Maxwell's theory. In this paper, we compare the two most widely used models of Maxwell's theory for superconductor AC loss simulations in analyzing ripple field losses: CSM and ECM. CSM is formulated in terms of the minimum magnetic energy variation (MMEV-formulation)~\cite{HacIacinphase} and ECM in terms of the $H$-formulation~\cite{brambilla07SST,hongZ06SST}. Both are formulated in 2D, considering only a transverse cross-section of an infinitely long superconductor in the axial direction. Hence, in our discussion, we define $E_z =: E$, $J_z =: J$, where $E_z$ and $J_z$ are the components of ${\bf E}$ and ${\bf J}$ in the axial direction of the conductors, respectively.

\subsection{Two models of the magnetoquasistatic Maxwell's theory: CSM and ECM}

\begin{figure}[!tb]
\centering
\includegraphics[width=8cm, height=8cm]{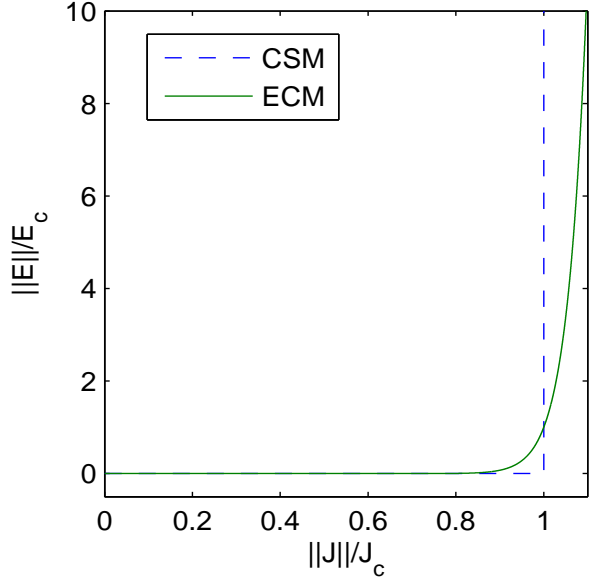}
\caption{The $E$-$J$-relations of CSM and ECM: the non-smooth relation of CSM (dashed line) and the power law used in ECM with $n$ = 25 (solid line).}
\label{fig:EJ-relations}
\end{figure}

When simulating AC losses of superconductors, the different models of the magnetoquasistatic version of the Maxwell's theory differ typically in the $E$-$J$-relation. Here, we are concerned with two models, CSM and ECM. In CSM, the $E$-$J$-relation is sharp: any $E$ in the superconductor will produce a $J$ in the direction of $E$, equal in magnitude to the local critical current density $J_{\mathrm{c}}$ in the wire. In ECM, $E$ and $J$ are still in the same direction but are related with the resistivity operator $\rho_{\mathrm{SC}}$, which can be expressed with the real number function
\begin{equation} \label{power-law}
\rho_{\mathrm{SC}} = \frac{E_\mathrm{c}}{J_\mathrm{c}}\left(\frac{||J||}{J_\mathrm{c}}\right)^{n-1},
\end{equation}
where $J_\mathrm{c}$ is the critical current density, generally a function of magnetic flux density ${\bf B}$ and temperature $T$, and $E_\mathrm{c}$ is the electric field criterion defining $J_\mathrm{c}$. In this article, we have used $E_\mathrm{c} = 10^{-4}$~V/m. The symbol $n$ denotes the $n$-value of the superconductor, related to vortex flux creep \cite{brandt95RPP} and macroscopically observed slanted shapes of the current-voltage curves, and $||J||$ denotes the Euclidean norm of the current density. In the limit $n \rightarrow \infty$ one obtains CSM from ECM. To summarize, the $E$-$J$-relations of CSM and ECM have been plotted in figure~\ref{fig:EJ-relations}.

As the $E$-$J$-relations used in CSM and ECM are different, the resulting descriptions of reality are different. The smooth $E$-$J$-relation of ECM leads to homogenization of the current density profile of the superconductor in low frequency or DC use; in the pure DC current case $J$ evolves towards constant distribution in which the local value of $J$ would be the total current divided by the cross-sectional area of the conductor. No such homogenization occurs in CSM. As we shall later show in ripple field problems, this difference manifests itself as different predictions in the loss behaviour, as well.

\subsection{The formulations}

As mentioned, the formulations used in our simulation codes are the MMEV-formulation of CSM and the $H$-formulation of ECM. We shall not go into the details of these well-established formulations here, but merely settle for a brief verbal overview. However, helpful references are given for convenience.

The variational principle of the MMEV-formulation, which minimizes the variation in magnetic energy to find the current distribution in the superconductor, was first introduced by Prigozhin \cite{prigozhin97IES}, and the formulation has later been developed gradually \cite{sanchez01PRB, HacIacinphase, souc09SST,pancakeFM}. The MMEV-formulation combined with a fast minimization method has been shown to be a very fast, efficient and reliable tool for AC loss simulations \cite{roebelcomp}.

The $H$-formulation, in which ECM is formulated entirely in terms of the magnetic field intensity ${\bf H}$ has quickly become the most popular formulation of ECM in the superconductor modelling community ~\cite{brambilla07SST,hongZ06SST, zhangM12SST, grilli13Cry, lahtinen12SST}. The popularity of the formulation is especially due to its good convergence properties, intuitivity of setting the driving quantities of the AC loss problem and the ease of implementation in commercial finite element method (FEM) software. The latter has been made possible by Comsol Multiphysics \cite{comsol} by including the basis functions needed for the FEM discretization of the $H$-formulation, time-stepping algorithms for solving stiff differential-algebraic equation (DAE) systems and the user interface for setting up a partial differential equation (PDE) of general form to be solved. The FEM discretization of the $H$-formulation is detailed in \cite{lahtinen12SST}.

\subsection{Computation of AC losses}

The total AC loss in a superconductor can be interpreted as pure ohmic loss \cite{acreview}. The ohmic power density $p_{\rmd}$ is defined as
\begin{equation} \label{PowerDensity}
p_{\rmd} = {\bf E} \cdot {\bf J} = EJ.
\end{equation}
The energy losses per unit length and cycle of AC field or current can then be found from
\begin{equation} \label{TotalLosses}
Q = \int_{T_\mathrm{p}} \left ( \int_S p_\mathrm{d} \rmd A \right) \rmd t,
\end{equation}
where $T_\mathrm{p}$ is the cycle of the AC quantity, $S \subset \Omega$ is the considered cross-section of the superconductor, and $t$ is time. In the case of pure AC situations $\int_S p_\mathrm{d} \rmd A $ should be integrated, for example, over the second cycle, to avoid the transient of the first one. However, as we shall see, this is not as straightforward in the case of DC-AC combinations.

When a superconductor carries a DC current, even clearly below its critical current $I_\mathrm{c}$, in an AC magnetic field, a DC voltage is produced over the superconducting sample if a certain threshold value $B^*$ for the AC field is exceeded \cite{ogasawara76Cry,mikitik01PRB}. The threshold value for a thin superconducting strip is \cite{mikitik01PRB}
\begin{equation} \label{Threshold}
B^* = \frac{\mu_0 J_{\mathrm{c}} d}  {2 \pi} \left[ \frac{1}{i} \ln{\frac{1+i}{1-i}} + \ln{\frac{1-i^2}{4i^2}}\right],
\end{equation} 
where $\mu_0$ is the vacuum permeability, $d$ is the thickness of the superconducting strip and $i$ = $I$/$I_{\mathrm{c}}$ is the ratio of the DC transport current to the critical current of the strip. In such a situation, part of the flux entering from another side of the sample is pumped through the sample and exits finally from the other side. This phenomenon is called dynamic magneto-resistance, and it is responsible for the produced voltage. As there is a voltage present in a current-carrying superconductor, losses associated to the transport current occur, too. In such cases, we can segregate the losses associated to magnetization of the superconductor, the \emph{magnetization losses}, and the losses associated to the transport current, the \emph{transport losses} \cite{acreview}.

\begin{center}
\begin{figure}[!tb]
%\centering
\includegraphics[scale=0.9]{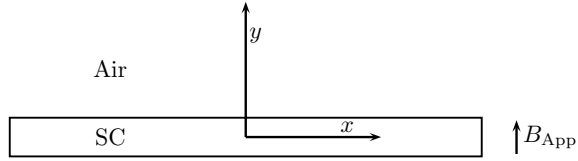}
\caption{The Cartesian $x$-$y$ coordinate frame attached to the center of the cross-section of the modelled superconducting (SC) sample. The thickness of the sample is highly exaggerated. The direction of the variation of the applied magnetic flux density $B_{\mathrm{App}}$ is also depicted; we consider only $y$-directional applied fields.}
\label{fig:CoordinateFrame}
\end{figure}
\end{center}

Magnetization loss can be computed as the area of the hysteresis loop, obtained by plotting the magnetization $M$ of the sample as a function of the applied field. In this article, we simulate structures with high aspect ratios and the field is applied perpendicular to the sample width. Assuming infinitely long conductors, we can identify $\Omega$ with a subset of the Euclidean real space $\mathbb{R}^2$ with the typical Cartesian metric and coordinates (see figure~\ref{fig:CoordinateFrame}). We define the magnetization contributing to the AC loss as
\begin{equation} \label{Magnetization}
M = \frac{1}{A_{\mathrm{SC}}}\int_{S} -xJ \rmd A,
\end{equation}
where $A_{\mathrm{SC}}$ is the area of the superconductor cross-section and the $x$-coordinate direction is perpendicular to the applied field~\cite{brandt93PRBa}. The magnetization loss $Q_{\mathrm{M}}$ in the steady-state is then computed from the area of the hysteresis loop. The loss is obtained as the integral
\begin{equation} \label{MagnetizationLoss}
Q_{\mathrm{M}} = A_{\mathrm{SC}} \oint -M(B_{\mathrm{App}}) \rmd B_{\mathrm{App}},
\end{equation}
where $B_{\mathrm{App}}$ is the $y$-component of the applied magnetic flux density. By using the symbol $\oint$ we emphasize that we are computing the area of the hysteresis loop obtained by plotting $M$ as a function of $B_{\mathrm{App}}$ over a full cycle of $B_{\mathrm{App}}$. However, if the steady-state has not been reached, the hysteresis loops do not close, and one cannot compute the magnetization losses from their areas. Having computed the total loss and the magnetization loss, one finds the transport loss $Q_{\mathrm{T}}$ simply as their difference:
\begin{equation} \label{TransportLoss}
Q_{\mathrm{T}} = Q - Q_{\mathrm{M}}.
\end{equation}

As a reference, we also use the analytical formula by Norris for calculating the hysteresis losses in a thin superconducting strip transporting an AC current, given by
\begin{equation} \label{NorrisLoss}
Q = \frac{\mu_0 {I_{\mathrm{c}}}^2} {\pi} \left[(1+i_\mathrm{m})\ln{(1+i_\mathrm{m})} + (1-i_\mathrm{m}) \ln{(1-i_\mathrm{m})} - {i_\mathrm{m}}^2 \right],
\end{equation}
where $i_\mathrm{m}$ is defined as $i$ before, but using the amplitude of the AC transport current, $I_m$, instead of DC current \cite{norris70JPD}.

\section{The methods of measurement}

The superconducting sample used in the measurements was a $Re$BCO coated conductor tape with cross-section dimensions 4 mm $\times$ 90 $\mu$m and a superconducting layer of approximately 4 mm $\times$ 1.4 $\mu$m \cite{SuperPower}. The properties of the tape, such as the $J_\mathrm{c}$-${\bf B}$- and $n$-${\bf B}$-dependencies, were well-known as they had been thoroughly investigated earlier \cite{pardo12SSTb}. We measured the current penetration into the sample carrying a DC current in zero external field by means of a Hall-probe mapping \cite{solovyov09IES, masti05MST}. Furthermore, we measured the AC losses of the sample carrying a DC current in an AC magnetic field perpendicular to the sample width using several different values of DC current and AC field.

\subsection{Current penetration measurement}

To investigate the penetration of current into the superconducting tape, we measured the component of the magnetic flux density perpendicular to the sample width right above the sample using the Hall-probe mapping technique. The measurement set-up is described in detail in \cite{solovyov09IES}. The Hall-probe was placed 0.3~mm above the sample. By means of an inversion procedure, we obtain the current distribution. The scans were made from $x = -$25~mm to 25~mm (figure~\ref{fig:CoordinateFrame}) and each scan took 181~s to complete.

\subsection{AC loss measurements}

All the AC loss measurements were performed by electrical means in liquid nitrogen (77~K). The contributions of the magnetization loss and transport loss were distinguished: the magnetization loss corresponds to the loss covered by the source supplying the AC magnetic field, and the transport loss corresponds to the loss covered by the source supplying the DC transport current \cite{ashworth99PhCa, vojenciak06SST}. The magnetization loss was measured using the calibration free method \cite{souc05SST} and the transport loss was obtained simply as the DC contribution of the voltage, V, over the superconducting sample multiplied with the DC transport current, $I_{DC}$. The voltage presents a high AC component (usually non-sinusoidal), being the waveform similar to a pulsed signal \cite{oomen99SST,uksusman09JAP}. However, this AC component does not contribute to the transport loss per cycle, $Q_{\rm tran}$, since $Q_{\rm tran}=\oint{\rm d}t V(t)I_{DC}=I_{DC}V_{DC}$, where $V_{DC}$ is the DC component of the voltage. The voltage was measured using a Keithley 2700 digital microvoltmeter. The AC component was suppressed by the built-in integrator at the analog/digital converter and additional digital filtering.

\begin{figure}[!tb]
\centering
\includegraphics[width=14cm]{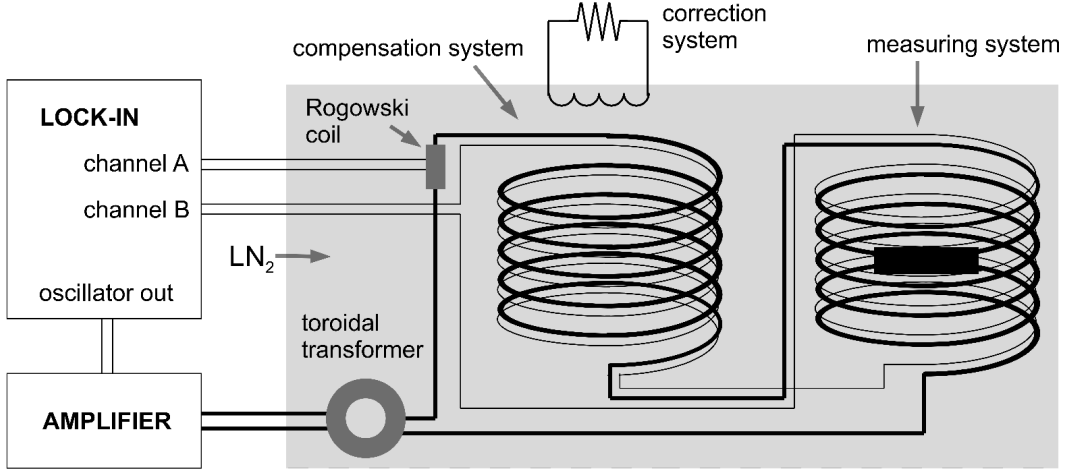}
\caption{Sketch of the calibration-free system to measure magnetization AC loss \cite{souc05SST}. The black rectangle represents the sample and the gray shaded area denotes the components immersed in liquid nitrogen. The correction system is magnetically coupled to the compensation system.}
\label{fig:sketch_CF}
\end{figure}

\begin{figure}[!tb]
\centering
\includegraphics[width=10cm]{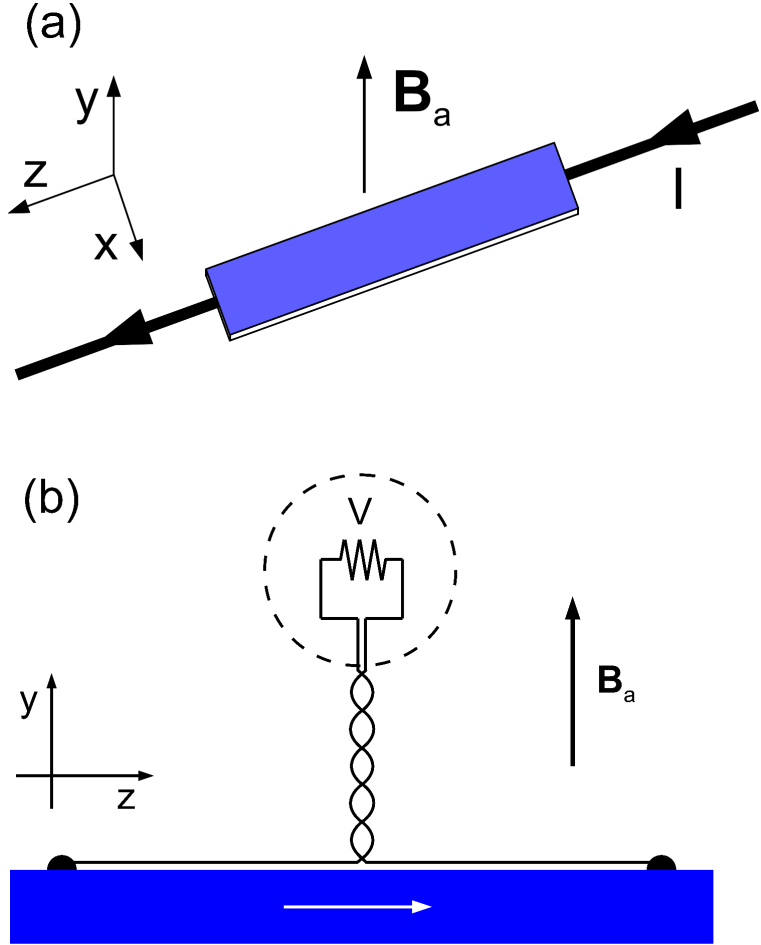}
\caption{(a) Simplified sketch of the measured situation. (b) Voltage taps for transport loss measurement. The sketch represents a side-view of the tape, where the solid rectangle is the superconducting tape, the semi-circles are the soldered terminals, the dashed line circular contour delimits the voltmeter with its equivalent circuit inside, and the white arrow shows the transport current direction. The wires are twisted from the tape surface to the voltmeter.}
\label{fig:sketch}
\end{figure}

\begin{figure}[!tbp]
\includegraphics[scale=0.5]{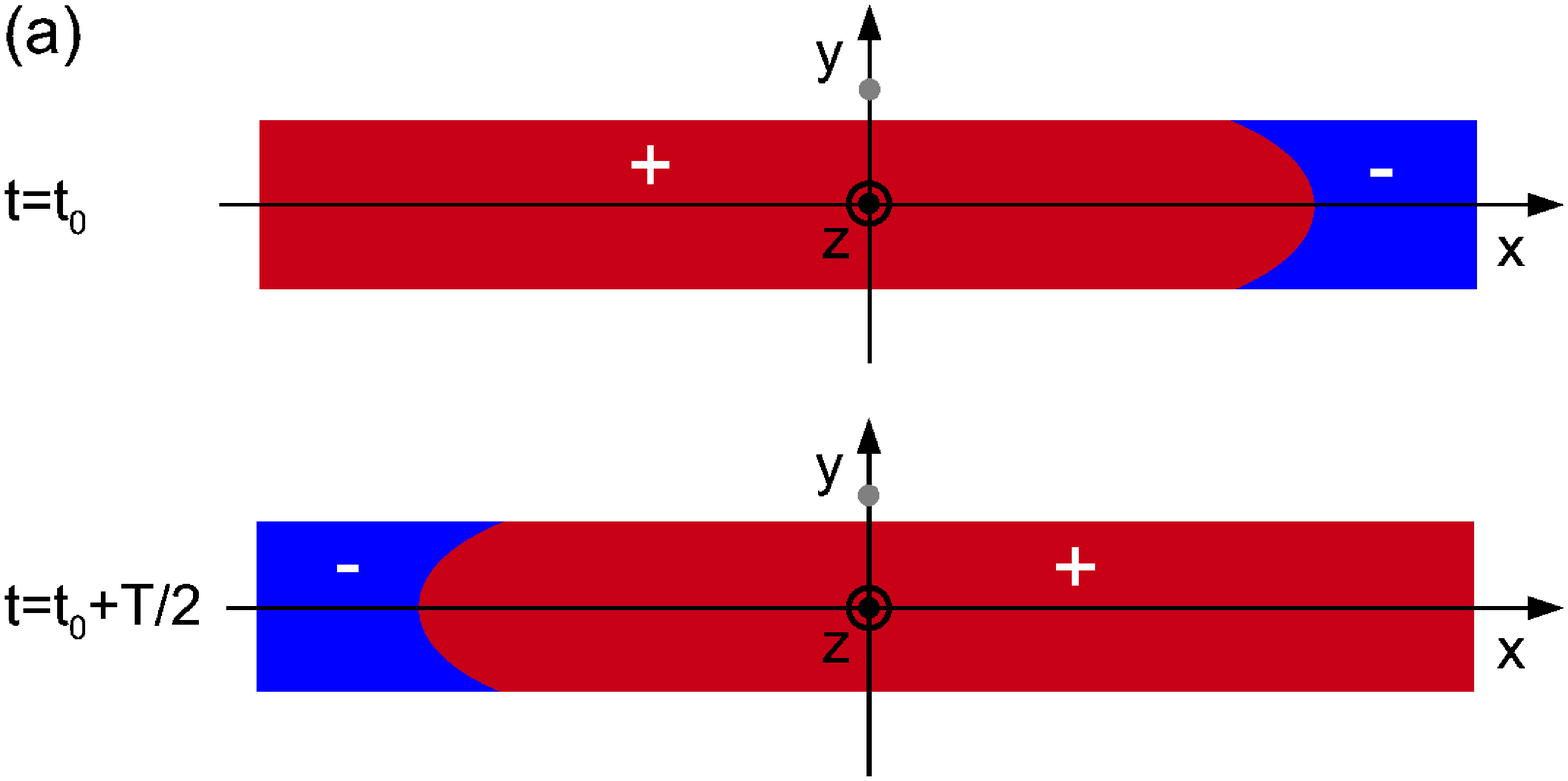}
\includegraphics[scale=0.5]{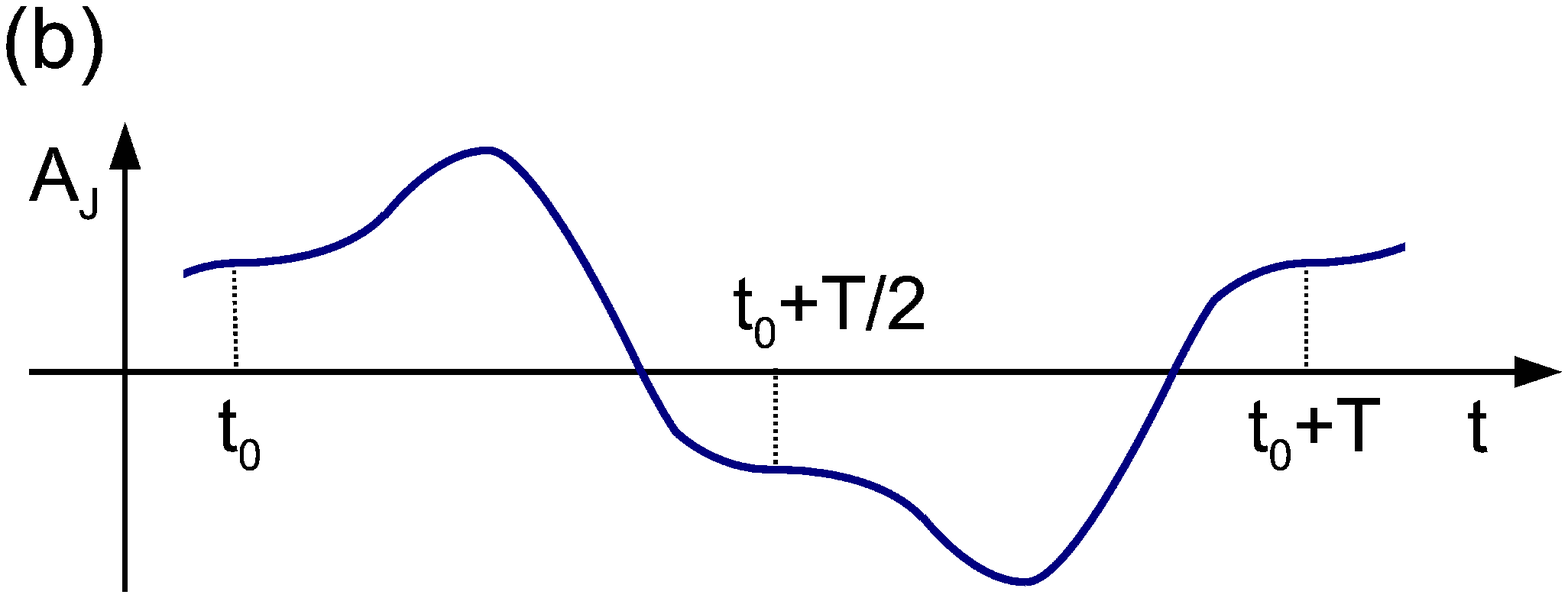}
\caption{(a) Sketch of the qualitative shape of the current density $J$ within the tape cross-section for an applied AC field in the $y$ direction while the tape transports DC current. The current density in the ``+'' and ``-'' regions is positive and negative, respectively, and its magnitude is not necessarily $J_c$. The current density follows the symmetry $J(x,y;t+T/2)=J(-x,y;t)$, where $T$ is the period of the AC field. (b) The current density above generates a vector potential $A_J$ on the $y$ axis which follows $A(x=0,y;t+T/2)=-A(x=0,y;t)$.}
\label{fig:sketchJA}
\end{figure}

In this work, we use a simple voltage-tap connection with the wire parallel to the tape length (figure \ref{fig:sketch}) in order to obtain the transport loss. This connection differs from the spiral-shaped or S-shaped loops previouly used in several articles \cite{ciszek02ACE,ciszek03PhC,duckworth11IES,rabbers98aPhC}. Although Spiral- and S-shaped loops are necessary for measuring the AC voltage that causes AC loss in AC current, simple voltage taps are sufficient to measure the DC voltage component in dynamic magneto-resistance experiments. The advantage of the latter connection type is that it is simple, compact, with low inductive signal, and suitable for windings. Although this kind of loop has already been successfully applied for dynamic magneto-resistance measurements \cite{rabbers98bPhC}, a theoretical explanation of its correcness has not been given. The explanation is the following. The equivalent circuit of the voltmeter is a resistance with a very high value (figure \ref{fig:sketch}). Then, the current in the connection wires is negligible and the electric field there is zero. Note as well that the wire thickness and the frequencies are low enough to neglect eddy currents in the wire. The electric field $\bf E$ in the connection wire is ${\bf E}=-\partial_t{\bf A}-\nabla \phi+\nabla \phi_c$, where $\bf A$ is the vector potential generated by the tape currents and applied magnetic field and $\phi$, $\phi_c$ are the scalar potentials generated by the charge densities in the tape and connection wire surfaces, respectively\cite{Note4}. Since the electric field in the connection wire vanishes, the voltage drop in the voltmeter, $V=\Delta\phi_c$, is
\begin{equation}
V=\Delta\phi-\int {\rm d}{\bf l}\cdot \partial_t{\bf A}, \label{Vloop}
\end{equation}
where the line integral is done over the length of the connection wire and ${\rm d}{\bf l}$ is the line differential. The voltage in the measuring loop from (\ref{Vloop}) is valid for any electromagnetic situation. For our case, since $\partial_t{\bf A}$ is periodic with no DC contributions, it vanishes after averaging $V$ in one cycle, and thus, it does not contribute to the DC voltage. This can be seen as follows. For our case, when the sample reaches the stationary state, the current density follows at least the symmetry described in figure \ref{fig:sketchJA}(a); which can be expressed as $J(x,y;t+T/2)=J(-x,y;t)$, where $T$ is the period of the AC excitation. Since in Coulomb's gauge the vector potential that $J$ creates is $A_J(x,y;t)=-{\mu_0}/({2\pi})\int_S{\rm d}S'\ln|{\bf r}-{\bf r'}|J(x',y';t)$, $A_J$ obbeys that $A_J(x,y;t+T/2)=-A_J(x,y;t)$. Therefore, for a pick-up coil with a straight segment located at $x=0$ (gray point in figure \ref{fig:sketchJA}(a)), the time dependence of $A_J$ is periodical and antisymmetric regarding a half cycle (figure \ref{fig:sketchJA}(b)), and hence its mean value (or DC component) vanishes. Since the applied AC magnetic field (and applied vector potential $A_a$) is also periodic and antisymmetric regarding a half cycle, the total $\partial_tA=\partial_t(A_J+A_a)$ presents no DC component. Then, from equation (\ref{Vloop}),
\begin{equation}
V_{DC}=\Delta\phi_{DC},
\end{equation}
and thence the DC component of the measured DC voltage corresponds to the DC component of the voltage supplied by the current source, given by $\Delta\phi$.  In addition, the measurement is independent on the shape of the connection wire circuit, as long as it is contained in the $yz$ plane of figure \ref{fig:sketchJA}(a), such as in our set-up (figure \ref{fig:sketch}).

Preliminary magnetization measurements showed an eddy-current contribution from the current leads. We removed this contribution as follows. A closed pick-up loop connected to a certain non-inductive resistance was placed in the compensation coil of the calibration-free system \cite{souc05SST}. The loss in the resistance can be adjusted by moving inwards and outwards the pick-up loop. In this way, we can remove an AC loss contribution from the measurements with the same frequency dependence as the eddy currents, that is, loss per cycle proportional to $f$. In the measurements, the loop position was adjusted in order to suppress the frequency dependence at no DC transport current. Since normal-conducting current leads are linear, adding a DC current does not change the eddy currents, and thus, in this way we remove the eddy current loss contribution for all DC currents.

\section{Results and discussion}

In this section, the simulation and measurement results are presented and discussed. The simulations were performed using home-made implementations of the MMEV-formulation and the $H$-formulation, programmed in Fortran and C++ programming languages, respectively. The FEM $H$-formulation solver has been implemented using the Riemannian geometry interface \cite{pellikka13CAM} implemented as a compact part of the open source Gmsh software \cite{gmsh}. For time-discretization of the $H$-formulation, the SUNDIALS package \cite{Sundials} has been used.

In the simulations that were compared with measurements, we used the ${\bf B}$-dependence of local $J_\mathrm{c}$ for the measured tape, obtained in \cite{pardo12SSTb}. For ECM, we also used the ${\bf B}$-dependence of the $n$-value, for which we did not take the angle of the field with respect to the tape into account but simply used the data for perpendicular applied field \cite{pardo12SSTb} as the local dependence of $n$ on ${\bf B}$. This is a good approximation when the applied field in the $y$-direction is present. However, in the pure net current case the $n$-value will be underestimated locally. In the AC loss simulations, the mesh contained either 1000 or 2000 triangular elements in the superconducting region for the ECM $H$-formulation, depending on the amplitude of the AC field, and for the CSM MMEV-formulation, we had 500, 1000 or 8000 rectangular elements in the superconductor. The meshes were dense enough for reliable simulations, as increasing the mesh density did not appreciably alter the results. We have a different number of elements when using the two formulations as they are substantially different: ECM in the $H$-formulation uses triangular elements and the degrees of freedom for $H$ are related to the edges of the mesh, while the CSM in MMEV-formulation uses rectangular elements with constant $J$. Some assumptions related to the heat generation are also different: for MMEV, the only elements that produce loss are those where $||J||=J_\mathrm{c}$ at least for part of the cycle, while for the $H$-formulation the loss generated in one element increases smoothly with $||J||$, generating loss also for $||J||<J_\mathrm{c}$; therefore modelling accurately the current density at the edges for low alternating currents or applied magnetic fields is even more important for the CSM. For simplicity, we used uniform meshes. For ECM, we divide the section in rectangular sections crossing the whole superconductor thickness with nodes at the corners and at the center, each rectangle containing 4 triangular elements. With the number of elements used, this results in rather high aspect ratio elements, which has been shown to yield good results when modelling high aspect ratio structures \cite{Zermeno}. For publications containing discussion about forming appropriate meshes for superconducting tapes, see, e.g., \cite{Zermeno, Morandi}. 

\subsection{Current penetration into a coated conductor tape}

To show the over-estimation of the homogenization of the current density profile in ECM, we measured the magnetic field generated by the coated conductor tape carrying a DC current using Hall-probe mapping. Furthermore, we simulated the same situation for the same tape.

After setting a DC current, the current density computed using Maxwell's theory evolves always to the distribution leading to minimal ohmic heat generation. If resistivity deviates from zero, the minimal total heat generation is achieved by a uniform current density profile in the cross-section of the wire. In superconductors, no heat is generated in pure DC use at currents substantially below $I_\mathrm{c}$, and thus, the current density profile will not converge towards the uniform one. Naturally, this is also what CSM predicts. However, the finite $n$-value used in ECM allows the homogenization of the current to happen: in ECM, there is always loss associated to a flowing current. No non-zero $J$ is associated to zero $E$, as $\rho$ connecting these two will never be strictly zero, whereas in CSM, only a \emph{changing} magnetic field can cause $E$ to deviate from zero. 

\begin{figure}[!tb]
\centering
\includegraphics[scale=0.9]{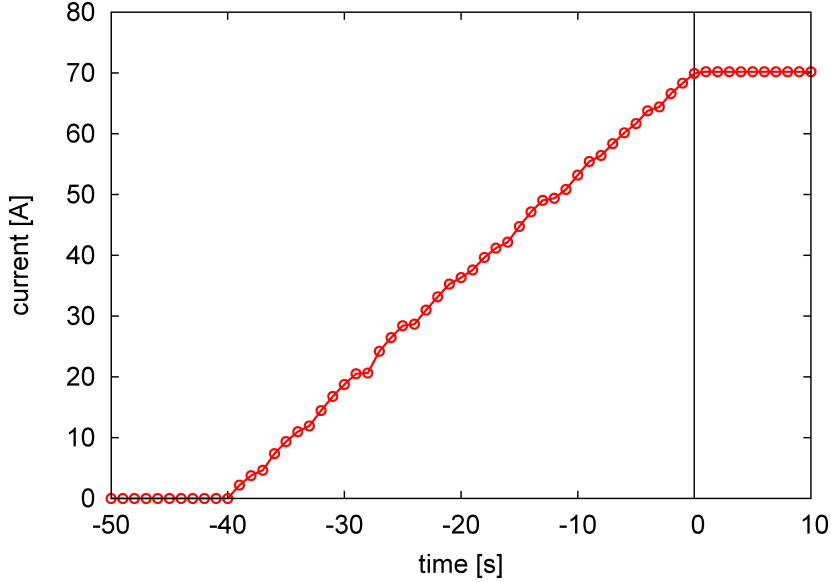}
\caption{The magnetic field maps were measured after setting a current of 70 A in a ramp increase. We take the time origin as the end of the ramp.}
\label{fig:ItMap}
\end{figure}

\begin{figure}[!tb]
\centering
\includegraphics[scale=0.9]{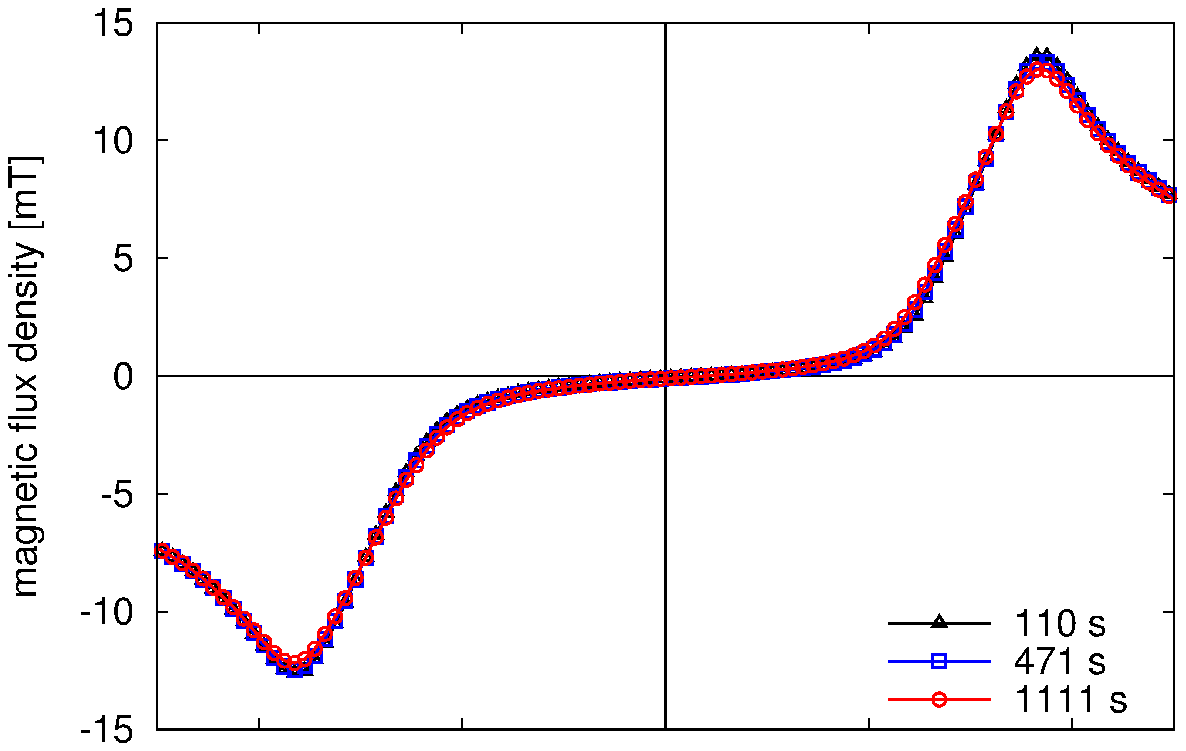}
\includegraphics[scale=0.9]{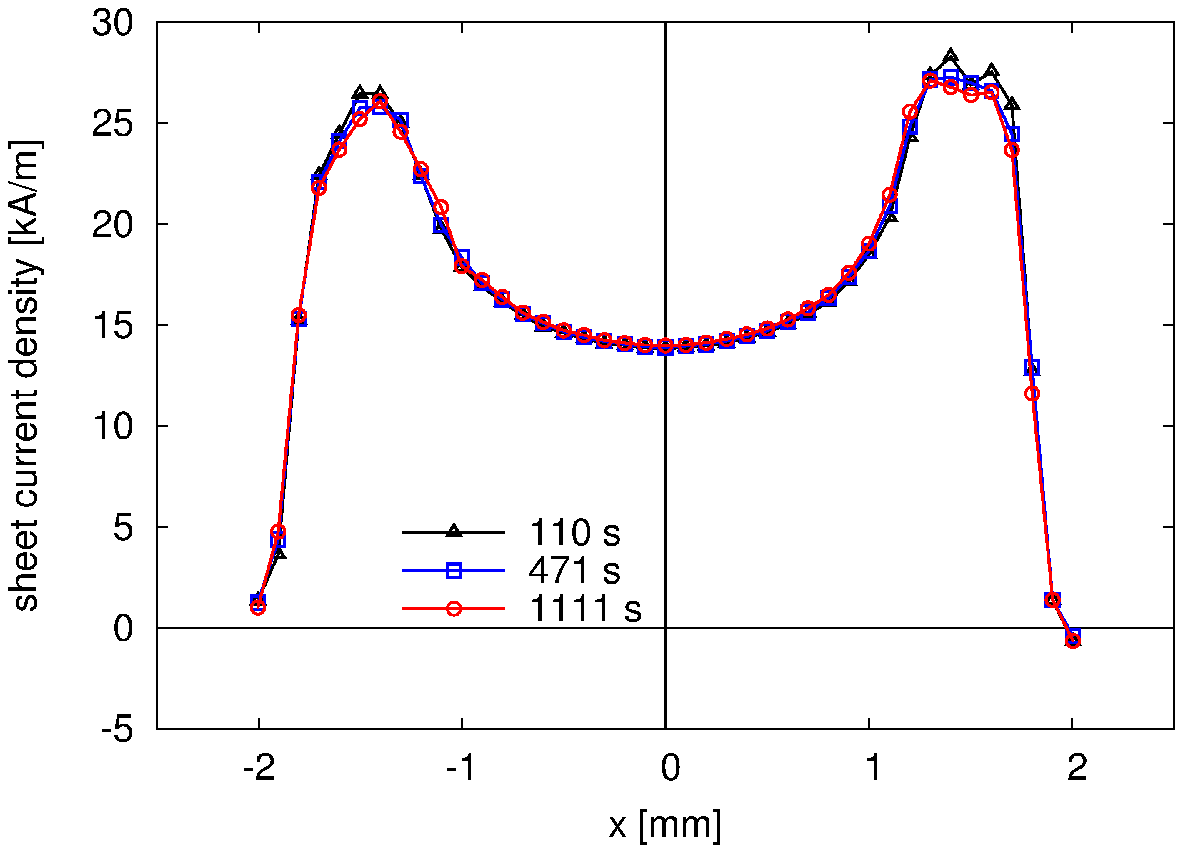}
\caption{Measured perpendicular component of the magnetic flux density (top) and sheet current density (bottom). The $x$-axis is defined in figure \ref{fig:CoordinateFrame}. The superconducting tape is located roughly from -2~mm to 2~mm on the $x$-axis. The legend indicates the time at which the field scan reaches the superconductor left edge in seconds.}
\label{fig:FieldMap}
\end{figure}

Figure~\ref{fig:FieldMap} depicts the measured magnetic field and obtained current density by inversion in the coated conductor tape carrying a DC current of 70~A after setting the current by the ramp increase of figure~\ref{fig:ItMap}. The self-field $I_\mathrm{c}$ for the tape is 128~A. Since the DC current is below $I_{\rm c}$, the sheet current density presents a wide sub-critical region between the peaks. The sheet current density corresponds to $J_{\rm c}d$ from the peaks to the edges. At the peaks, the magnetic field vanishes (more precisely, the average magnetic field over the tape thickness), and hence $J_c$ is maximum. One observes that the field generated by the current, and thus the current density profile itself, remains \emph{exactly the same} within the measurement accuracy at least for the first 15 minutes.

\begin{figure}[!htb]
\centering
\includegraphics[scale=0.9]{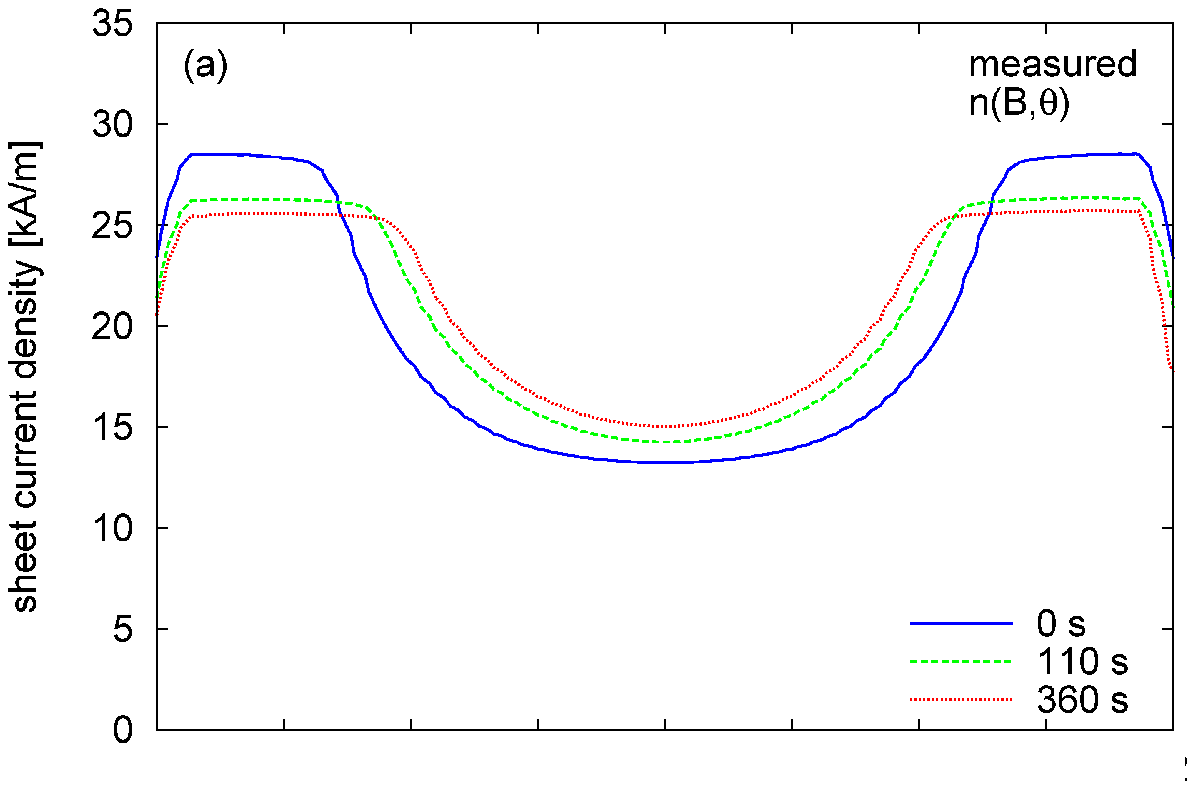}
\includegraphics[scale=0.9]{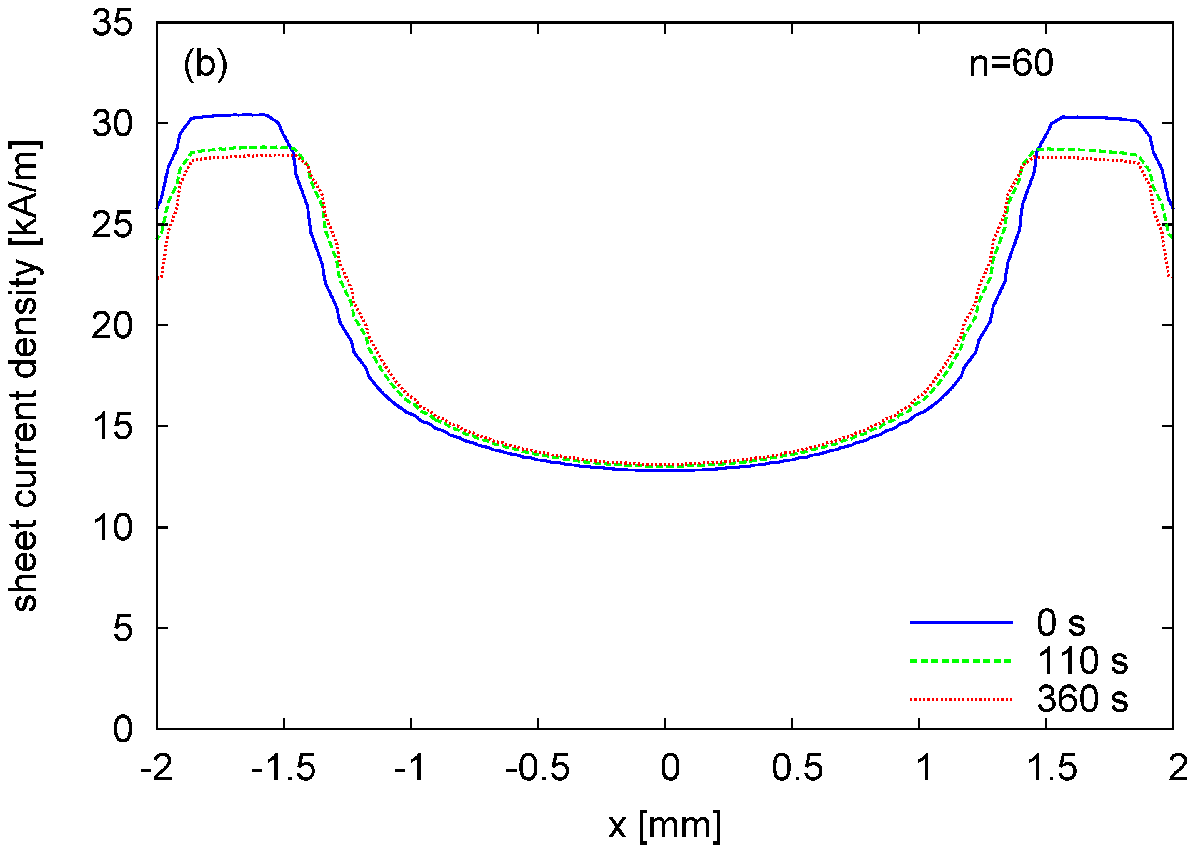}
\caption{The one-dimensional profiles of current density as a function of the tape width (tape center at $x$ = 0) as predicted by ECM at $t$ = 0~s, $t$ = 110~s and $t$ = 360~s plotted on a line across the tape for both the measured $n$-${\bf B}$ dependence (a) and $n=60$ (b); both cases use the $J_c$-${\bf B}$ dependence extracted from measurements. Note that in the simulation, the current penetration was not purely one-dimensional, as there were two to three elements in the tape thickness throughout the tape.}
\label{fig:Profiles1D}
\end{figure}

In comparison, we simulated the same situation using ECM: the current was raised in 40~s up to its amplitude value 70~A and kept constant up to 360~s. The one-dimensional current density profiles in the width of the tape in such a case are depicted in figure~\ref{fig:Profiles1D}. Here, the tendency to converge towards a uniform current distribution is evident already in few seconds, and the qualitative behaviour of the profile is remarkably different from the measurement result. In ECM, the further the time evolves, the more homogeneous is the current density profile, whereas nothing like this is observed in the measurement. Hence, the power law used in ECM as the \emph{local} $E$-$J$-relation leads here to non-physical behaviour: a relation closer to CSM would yield a better description of reality in such a case. In order to check this, we calculated the same situation with a much larger $n$ value ($n=60$) and have observed a much slower relaxation. Although for our experimental $n$, we did not correct the self-field, the actual $n$ cannot reach values as large as 60, at least close to the tape edges, where the magnetic field is of the order of 20~mT or larger. Hence the result does indeed show that the homogenization is over-estimated. Naturally, however, further research on the topic is required.

\subsection{The frequency dependence of hysteresis losses predicted by ECM}

As observed, in DC / low-frequency simulations ECM leads to very different predictions than CSM, and in terms of current penetration, its predictions differ from measurements, as well. Now, we will show through simulations, that the hysteresis losses per cycle of AC current, as predicted by ECM, also depend on the frequency. While there are measurements supporting the assumption of them being rather well frequency-independent, at least for currents substantially below $I_\mathrm{c}$ \cite{ciszek96SST, herrmann00PhC}, some experimental data for high-temperature superconductors do suggest a frequency-dependence of hysteresis losses \cite{polak07SST, wakudaCryog97, eckelmanPhC98, thakur111, thakur112}, especially close to or above $I_\mathrm{c}$ or full field penetration. Even though the frequency-dependence of the AC loss predictions of ECM has been investigated earlier as well \cite{amemiya97IES, thakur111, thakur112, sander10JoP}, we include such simulations for completeness and to support our other observations. In \cite{yamafujiCryog97}, it was suggested that the power law based ECM does not properly describe the frequency-dependence of AC losses in HTS, but a \emph{generalized critical state model} yields a better description of reality. This complements our observation of ECM not describing the DC behaviour of the coated conductor tape well enough, as detailed in the previous subsection.

\begin{figure}[!tb]
\centering
\includegraphics[scale=1.0]{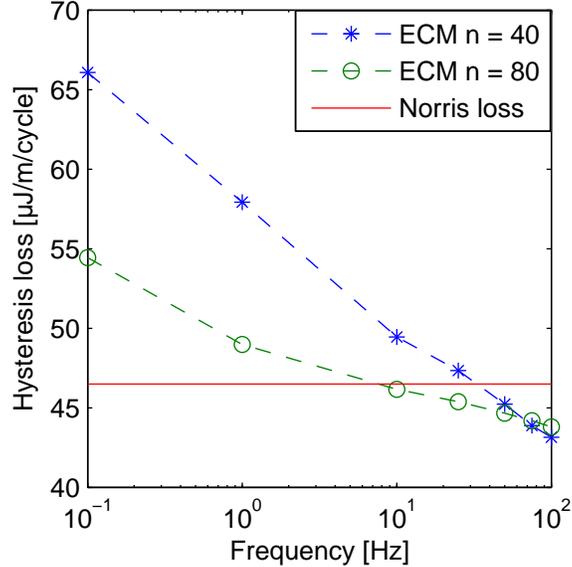}
\caption{Hysteresis losses predicted by ECM for AC transport current of $0.5I_\mathrm{c} = 50$~A at several frequencies. Norris loss for this case is 46.496~$\mu$J/m/cycle.}
\label{fig:ECMFreq}
\end{figure}

Using ECM based $H$-formulation, we simulated a superconducting tape with cross-section dimensions 4~mm~$\times$~1~$\mu$m and $I_\mathrm{c} = 100$~A transporting a sinusoidal AC current of amplitude 0.5$I_\mathrm{c}$ at several frequencies from 0.1 to 100~Hz. We used two $n$-values, $n = 40$ and $n = 80$. The results are displayed in figure~\ref{fig:ECMFreq}. The frequency-dependence of the results is very clear, especially for $n = 40$. For a fixed $n$-value, the lower the frequency, the higher the losses. The measurements, e.g., in \cite{polak07SST, thakur112} do support this kind of behaviour, but not to this extent, even though the discrepancy is not enormous. The frequency-dependence of the hysteresis losses is reduced, as the model is modified towards CSM by increasing the $n$-value to 80. Naturally, the losses obtained from CSM based calculations would not depend on the frequency. Comparing the results with the ones obtained using the analytical formula \eqref{NorrisLoss}, based on CSM, we see that at the frequencies of power applications, ECM gives rather similar loss estimates. At low frequencies, however, ECM over-estimates the hysteresis losses. This over-estimation is caused by the over-estimation of the homogenization of the current distribution. The loss in ECM constitutes of two compononents: the stationary loss, which is related to the dissipation in the stationary state, and the loss related to time-varying magnetic field, the variation loss. At low frequencies, the relative contribution of the stationary loss over a cycle of AC field is large, whereas at high frequencies, it is small. As the movement of the current front in a DC/low-frequency current case is over-estimated by ECM, so should be the loss in such a situation: homogenization of the current density profile is associated to the tendency of the stationary loss to minimize over time. Using a local $E$-$J$-relation closer to CSM, suggested by the current penetration measurement of the previous section, would lead to smaller frequency-dependence.

\subsection{Comparison of the models: ripple field loss simulations with constant $I_{\mathrm{c}}$}

To make a thorough comparison of the models in ripple field loss simulations, we performed a set of simulations with different combinations of DC field, DC current and AC field for a coated conductor tape with cross-section dimensions 4~mm~$\times$~1~$\mu$m. To make the interpretation of the results easier, we used a constant critical current density, yielding a constant critical current $I_\mathrm{c}$~=~100 A, and the $n$-value used in the power law of ECM was 40. These simulations were not compared with measurements, but were only performed to compare the predictions of the models. We used low-amplitude AC fields and a DC current clearly below $I_\mathrm{c}$ as in such situations the properties of the models are well exposed.

\subsubsection{AC losses per cycle}

\begin{figure}[!tb]
\centering
\includegraphics[width=8cm]{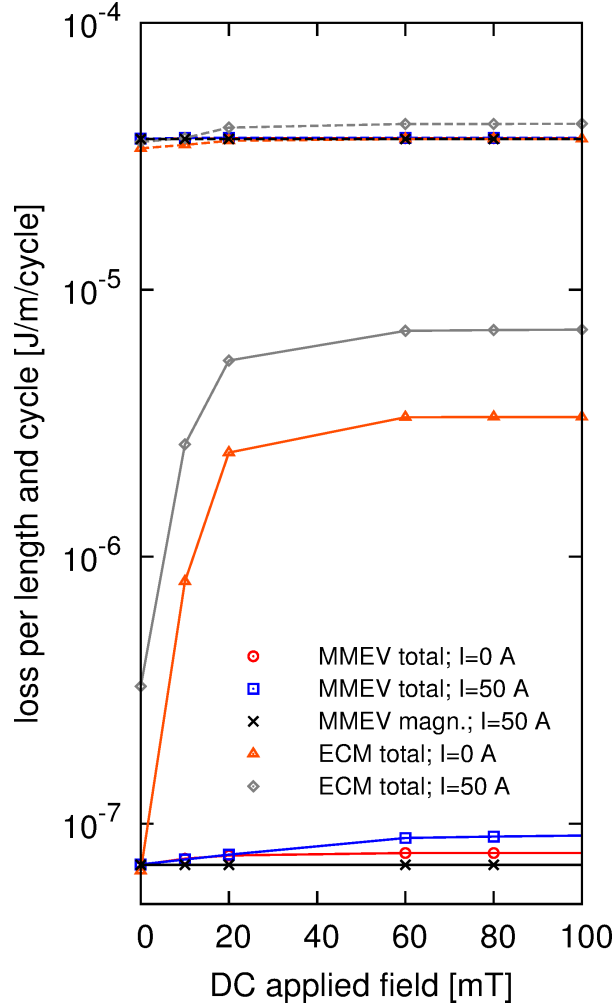}
\caption{Simulated total and magnetization loss from table \ref{table:Losses}, corresponding to a 4mm$\times 1\mu$m tape with constant $J_c$ and $I_c=$100 A. The continuous and dashed lines are for AC applied field amplitudes of 1 and 5 mT, respectively. The symbols (and colours) identify both continuous and dashed curves. The magnetization loss for MMEV at $I$=50 A is identical to that at $I$=0 A, and thus, it is not shown.}
\label{fig:compBean}
\end{figure}

As the AC ripple field we used a sinusoidal function $B_{\mathrm{App}} = B_{\mathrm{a}}\sin (2 \pi f t)$ with frequency $f$~=~50~Hz and a varying amplitude $B_{\mathrm{a}}$. In each case with both DC and AC quantities involved, the DC quantities were raised up to their target values simultaneously in 5~ms, and after keeping them constant for 15~ms, the AC ripple field was applied on top of them. The problem was then integrated in time for 10 cycles of the AC field (200~ms), and the AC losses were integrated over the last cycle. The AC losses per cycle in the tape are presented in table~\ref{table:Losses} and figure \ref{fig:compBean}. The slight increase in the total loss for MMEV at zero DC current is a numerical artifact, due to discretization of the superconductor cross-section. However, the results for $I$=50 A are qualitatively correct.

\begin{table}[!tb]
\begin{ruledtabular}
\caption{The AC losses in the superconductor for different values of DC field, DC current and AC field computed using the CSM based MMEV-formulation and the ECM based $H$-formulation. The AC fields are given as their amplitude values $B_{\mathrm{a}}$.}
\label{table:Losses}
\scriptsize 
    \begin{tabular}{ p{1.2cm}  p{1.2cm}  p{0.5cm}  p{1.8cm}  p{1.5cm} p{1.8cm} }  % p{1.5cm} }     \br
    DC field [mT] & DC current [A] & AC field [mT] & Total loss (CSM) [J/m/cycle] & Total loss (ECM) / Total Loss (CSM) & Magnetization loss (CSM) [J/m/cycle]  \\  %&  Magn. loss (ECM) / Magn. loss (CSM) \\ \mr
    \multirow{4}{*}{0}  				& \multirow{2}{*}{0}    & 1 	& $ 7.013 \times 10^{-8}$  & $ 0.949 $ & $ 7.012 \times 10^{-8} $ \\ % & $ $ 					\\	
    															&  & 5   &										 $ 3.677 \times 10^{-5}$ & $ 0.922  $ 	& $ 3.677 \times 10^{-5} $ \\ % & $  $						 \\
    														& \multirow{2}{*}{50}   & 1   & $ 7.036 \times 10^{-8}$  & $ 4.644  $   & $ 7.012 \times 10^{-8} $ \\ % &  $ - $					\\	
    															&  & 5      							& $ 3.680 \times 10^{-5}$  &   $ 0.972  $   & $ 3.677 \times 10^{-5} $ \\ % &  $  $						\\
    \hline
    \multirow{4}{*}{10}  				& \multirow{2}{*}{0}    & 1 	& $ 7.397 \times 10^{-8}$  &$ 				10.930 $  & $ 7.012 \times 10^{-8} $  \\ % & $ - $ 						\\	
    															&  & 5   & 								$ 3.690 \times 10^{-5}$  & 		$  0.947 $ 	& $ 3.677 \times 10^{-5} $ \\ %  & $ $						 \\
    														& \multirow{2}{*}{50}   & 1 & $ 7.344 \times 10^{-8} $  & $ 				35.849 $    & $ 7.019 \times 10^{-8} $ \\ % &  $ - $					\\	
    															&  & 5      							& $ 3.704 \times 10^{-5}$  & 	$ 1.000 $ & 	$ 3.677 \times 10^{-5} $ \\ % & $ $ \\
    \hline
    \multirow{4}{*}{20}  				& \multirow{2}{*}{0}    & 1 	& $ 7.601 \times 10^{-8}$  & $ 				32.29 $  & $ 7.012 \times 10^{-8} $ \\ % & $ - $ 						\\	
    															&  & 5   									& $ 3.696 \times 10^{-5}$  & 	$ 0.979 $  	& $ 3.677 \times 10^{-5} $ \\ % & $  $						 \\
    														& \multirow{2}{*}{50}   & 1 & $ 7.662 \times 10^{-8}$  & 	$ 				 70.937 $   & $ 7.018 \times 10^{-8} $ \\ %& $ - $					\\	
    															&  & 5      							& $ 3.707 \times 10^{-5}$  & 	$  1.094	$   & $ 3.677 \times 10^{-5} $ \\ % & $ $ 				\\
    \hline
    \multirow{4}{*}{60}  				& \multirow{2}{*}{0}    & 1 	& $7.761 \times 10^{-8}$  & $				42.935 $ & $ 7.012 \times 10^{-8} $ \\ % & $ - $  						\\	
    															&  & 5   & $3.700 \times 10^{-5}$  &										$ 0.993 $ & $ 3.677 \times 10^{-5} $ \\ % &	$ $						 \\
    														& \multirow{2}{*}{50}   & 1	& $8.841 \times 10^{-8}$  & 	$ 				79.387	$ & $ 7.019 \times 10^{-8} $ \\ % & $ - $  					\\	
    															&  & 5      							& $3.712 \times 10^{-5}$  &	$ 1.128 $ & $ 3.677 \times 10^{-5} $ \\ % &  $ $ \\
    \hline
    \multirow{4}{*}{80}  				& \multirow{2}{*}{0}    & 1 	& $7.763 \times 10^{-8}$  & $  			42.996 $ & $ 7.012 \times 10^{-8} $ \\ % &  $ - $ 						\\	
    															&  & 5   & $3.700 \times 10^{-5}$  & 									$ 0.993 $ & $ 3.677 \times 10^{-5} $ \\ % & $ $							 \\
    														& \multirow{2}{*}{50}   & 1  & $8.951 \times 10^{-8}$  & $				  78.966 $ & $ 7.019 \times 10^{-8} $ \\ % & $ - $  					\\	
    															&  & 5      							& $3.712 \times 10^{-5}$  &  $ 1.128 $ & $ 3.677 \times 10^{-5} $ \\ % & $ $ \\
    \hline
    \multirow{4}{*}{100}  			& \multirow{2}{*}{0}    & 1 	& $7.763 \times 10^{-8}$ & $					43.002 $ & $ 7.012 \times 10^{-8} $ \\ % & $ - $  						\\	
    															&  & 5   & 									$3.700 \times 10^{-5}$  &  $ 0.993$ & $ 3.677 \times 10^{-5} $  \\ % & $ $ 							 \\
    														& \multirow{2}{*}{50}   & 1 & $9.022 \times 10^{-8}$  & 	$ 				78.590 $ & $ 7.019 \times 10^{-8} $  \\ % & $ - $   					\\	
    															&  & 5      							& $3.712 \times 10^{-5}$  & 	$ 1.128 $ & $ 3.677 \times 10^{-5} $	 \\ % & $  $ \\

    \end{tabular}
\end{ruledtabular}
\end{table}

\begin{figure}[!tb]
\centering
\subfigure[]{
\includegraphics[height=7cm]{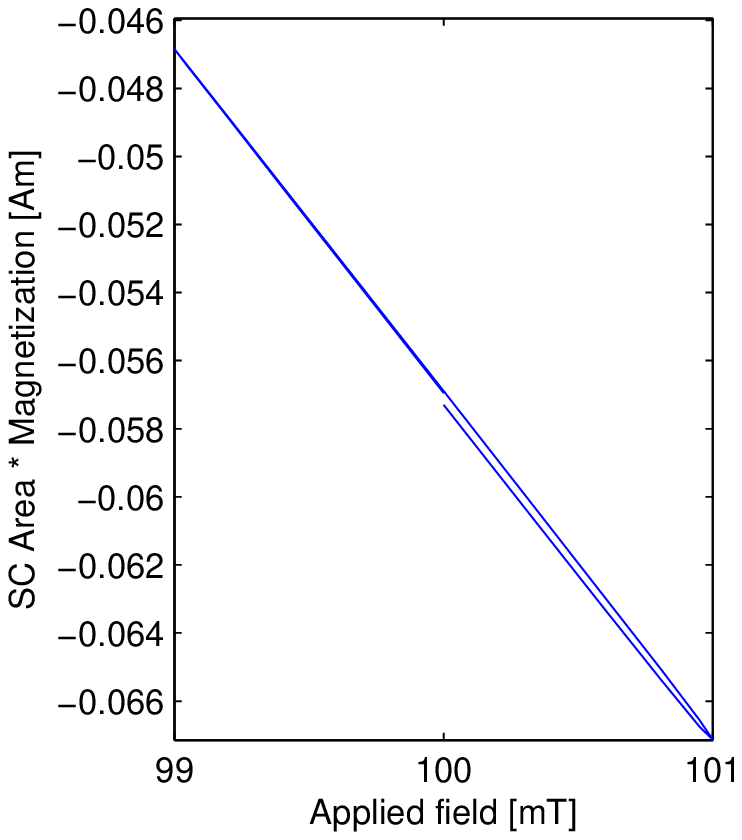}
\label{fig:loop1}
}
\subfigure[]{
\includegraphics[height=7cm]{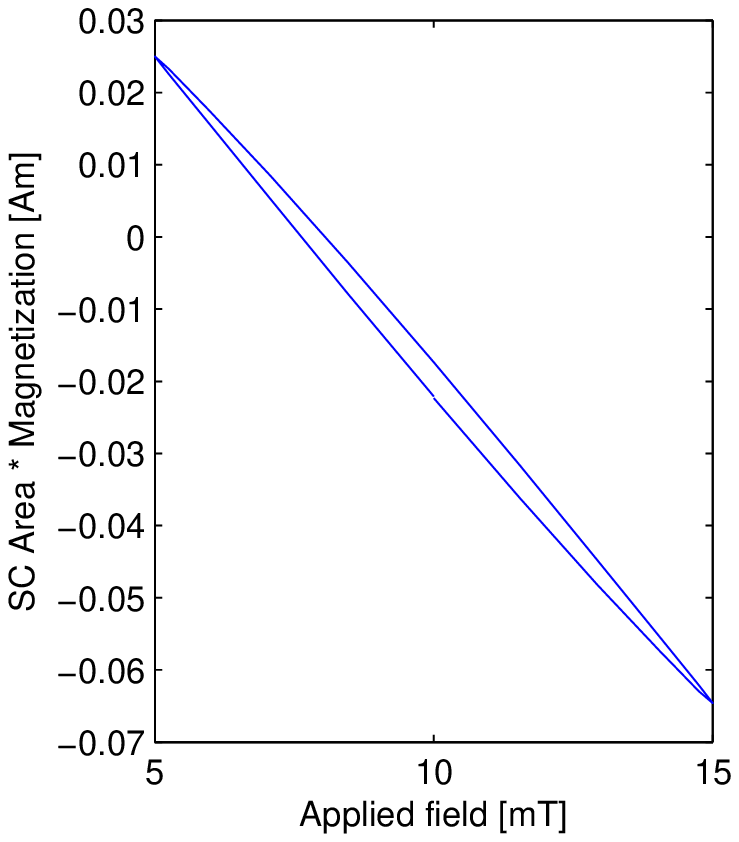}
\label{fig:loop2}
}
\caption{Hysteresis loops over the 10th cycle of the AC ripple field obtained using ECM for cases of (a) 100~mT DC field, 50~A DC current and 1 mT AC field and (b) 10~mT DC field, 50~A DC current and 5~mT AC field. In the latter case, the DC field is small and comparable to the AC field, and thus, the loop is almost closed, whereas in the former case, it is obvious that the loop is still far from a closed one.}
\label{fig:loops}
\end{figure}

In the pure AC cases (neither DC applied field nor DC current), the two models yield rather similar predictions. However, as soon as there is a DC current involved, there is a large discrepancy in the predictions for very low AC fields, 1~mT. As also the DC field is added, the difference further increases. For the 5~mT cases, the predictions of the models are in reasonable agreement for all the cases. Again, ECM over-estimates the loss related to the low-frequency quantities. The distribution of loss between magnetization and transport losses was only investigated for CSM based simulations, as for ECM, the steady-state was not reached after 10 cycles and the hysteresis loops thus did not close, if DC field was involved (figure \ref{fig:loops}). This was especially prominent with low AC fields, 1~mT. Using CSM, the loops closed to a reasonable extent for all the cases. 

\subsubsection{The dissipated power as a function of time}
\label{s.power}

The finite $n$-value leading to over-estimation of the homogenization of current density in ECM is related to the discrepancies in the loss behaviour between the predictions of the MMEV-formulation and the $H$-formulation in very low AC fields of 1~mT amplitude. This may be further investigated by plotting the dissipated power $P$ per unit length of the superconductor as a function of time $t$. For the 1~mT cases, there is a significant, slowly descending offset in the $P(t)$ curves obtained using ECM.

\begin{figure}[!tb]
\centering
\includegraphics[scale=1.0]{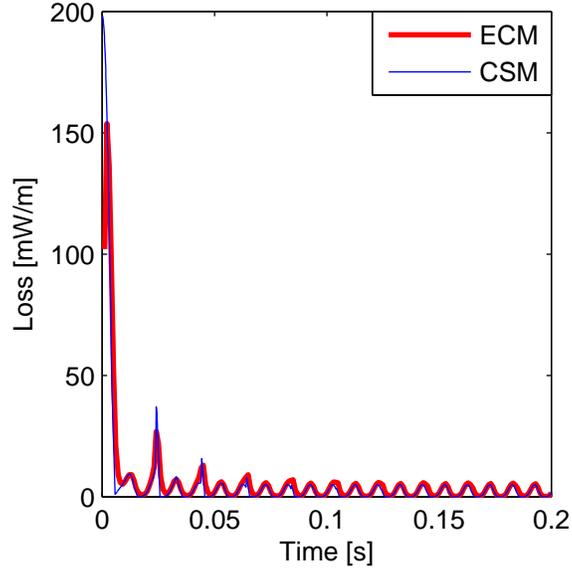}
\caption{The simulated $P(t)$ curves for 50~A DC current, 100~mT DC field and 5~mT AC ripple field.}
\label{fig:Pt_5mT}
\end{figure}

Let us first investigate a case, in which the predictions of the models agree with each other. In figure~\ref{fig:Pt_5mT} the $P(t)$ curves for a case of 5~mT AC Field, 100~mT DC field and 50~A DC current are depicted. The curves are obviously very similar both in shape and in magnitude. The descending offset due to current density homogenization in ECM based curve is not significant compared to the AC field, and thus the $P(t)$ curves of the models are similar. In the curve obtained using CSM, the first few peaks are somewhat higher than in the one obtained using ECM, but soon both of them settle to constant level of approximately 5~mW/m. The difference in the first peaks can be partially explained by a higher sampling frequency used in the CSM based MMEV solver. 

\begin{figure}[!tb]
\centering
\includegraphics[scale=1.0]{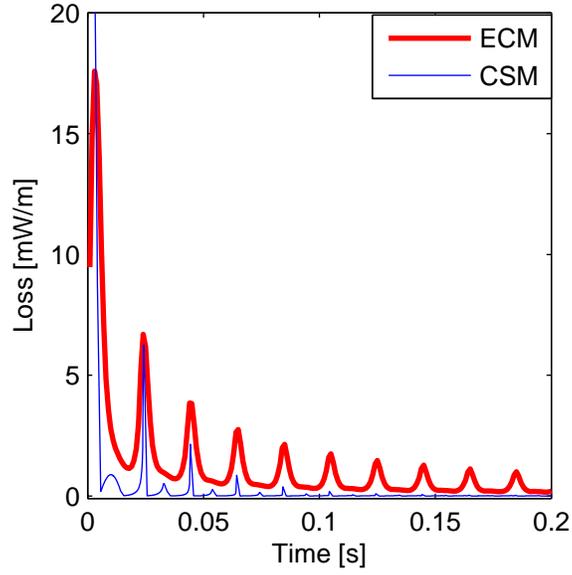}
\caption{The simulated $P(t)$ curves for 50~A DC current, 100~mT DC field and 1~mT AC ripple field. The ripple field starts at $t = 0$. At that time, the DC quantities have already been raised to their target values in 5~ms and kept constant for 15~ms.}
\label{fig:Pt_1mT}
\end{figure}

Figure \ref{fig:Pt_1mT} shows the $P(t)$ curves obtained using CSM and ECM for a case of 1~mT AC field, 100~mT DC field and 50~A DC current. As one observes, now both the shapes of the $P(t)$ curves and the orders of magnitude of the losses are very different. Because of the slow homogenization of the current density profile in ECM, there is a slowly descending offset in the $P(t)$ curve computed with the $H$-formulation based solver, not to be seen in the results obtained with the MMEV-formulation. This descending offset is related to the DC components of the current and magnetic field. The loss will not be fully stabilized until the current density profile is completely homogeneous, which would take hundreds or thousands of ripple field cycles. This data thus supports our observation, that ECM over-estimates the losses associated to DC quantities. Also in both curves, the maximum level of the peaks keeps decreasing still after 10 cycles of the ripple field.

\begin{figure}[!tb]
\centering
\subfigure[]{
\includegraphics[scale=0.9]{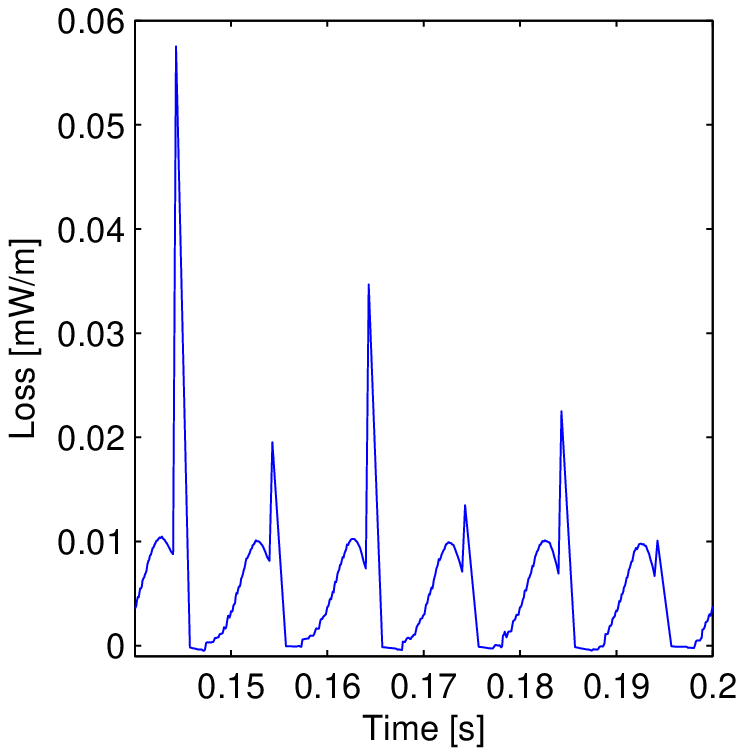}
\label{fig:Pt_1mT_ECM_closeup}
}
\subfigure[]{
\includegraphics[scale=0.9]{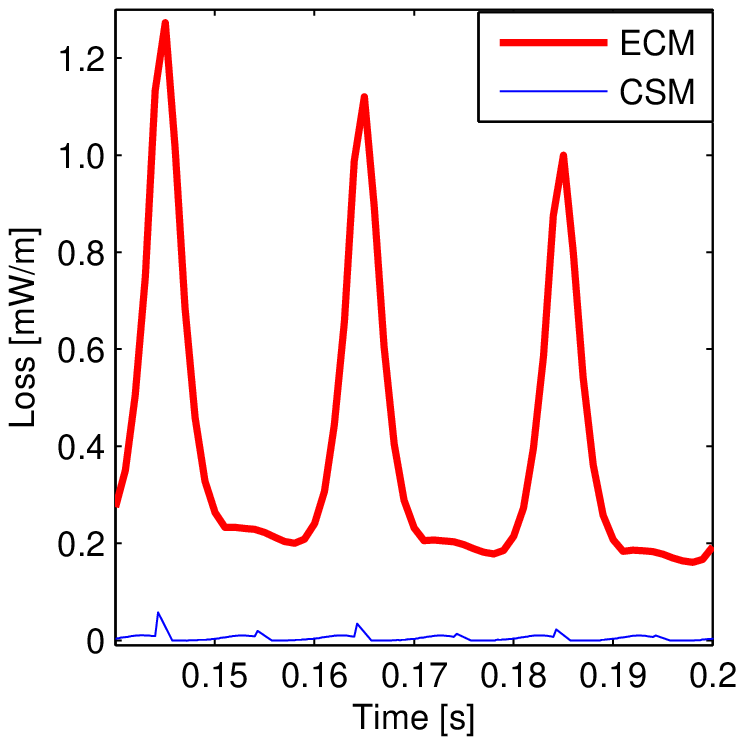}
\label{fig:Pt_1mT_CSM_closeup}
}
\caption{Close-ups of the $P(t)$ curves of figure~\ref{fig:Pt_1mT} from 0.14~s to 0.20~s, obtained using (a) CSM based MMEV-formulation and (b) ECM based $H$-formulation. In (b), also the curve of (a) has been plotted for direct comparison of the results.}
\label{fig:Pt_1mT_closeup}
\end{figure}

\begin{figure}[!tb]
\centering
\subfigure[]{
\includegraphics[scale=0.9]{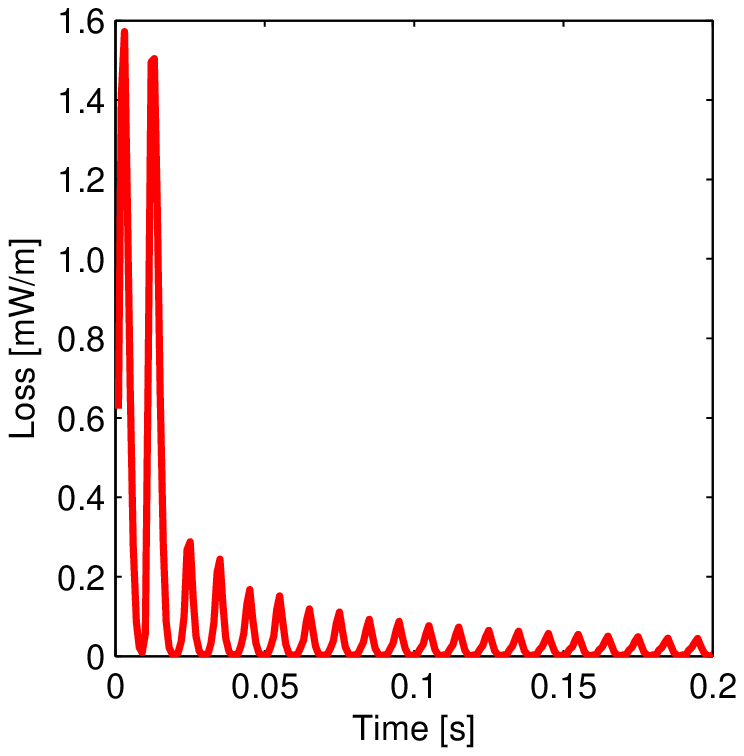}
\label{fig:Pt_0mT_50A_1mT_ECM}
}
\subfigure[]{
\includegraphics[scale=0.9]{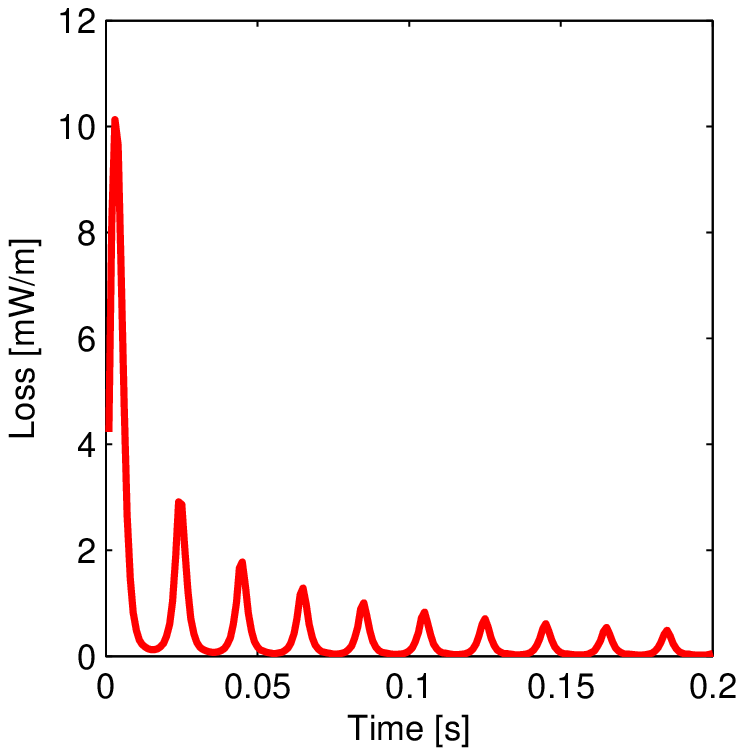}
\label{fig:Pt_10mT_50A_1mT_ECM}
}
\caption{$P(t)$ curve for (a) 50~A DC current, 0~mT DC field and 1~mT AC ripple field exhibiting two peaks per cycle and (b) 50~A DC current, 10~mT DC field and 1~mT AC ripple field exhibiting one peak per cycle, obtained from simulations using ECM based $H$-formulation.}
\label{fig:ECM_0_DC_FIELD}
\end{figure}

In figure~\ref{fig:Pt_1mT_closeup}, a close-up of the curves near 0.2~s is depicted. Note, that even if the bases of the peaks of the ECM based $P(t)$ curve were at zero, the order of magnitude of the losses would still be higher than for CSM. In both curves, the rising half-cycle of the AC field creates a notably higher peak than the descending one. In the curve obtained using ECM, the peak of the descending half is barely visible. However, in the CSM based computations, the difference is not as dramatic. Also, the CSM based $P(t)$ curve exhibits a very sharp peak near the end of each half cycle, not to be seen in the ECM based curve. These secondary peaks are due to a progressive current penetration in the boundaries between $J=J_\mathrm{c}$ and $J=-J_\mathrm{c}$ caused by the alternating excitation, similar to those in tapes subject to a DC applied magnetic field and a small perpendicular AC component \cite{mikitik04PRBb}. In addition, looking at figure~\ref{fig:ECM_0_DC_FIELD}, one observes that if there is no external DC field involved, but only a DC current flowing through the conductor, the ECM based $P(t)$ curve does have two visible peaks per cycle, but as soon as the external DC field is added, the second peak will be negligible in comparison with the first one.

\subsubsection{Concluding remarks}

To sum up this comparison, the models give very similar results for a wide range of ripple field cases. However, at very low AC fields of 1~mT amplitude with significant DC fields or currents, the over-estimation of the loss related to DC quantities exhibited by ECM leads to discrepancies in the loss predictions. This is a consequence of the difference in the $E$-$J$-relations of the models. The cases with 1~mT AC field amplitude also exhibit very long transients observed in the $P(t)$ curves, for both ECM and CSM. This suggests that the transients are partially due to complex flux-front patterns, which will be discussed in detail in our future work. In ECM, they are partially due to the slowly decreasing stationary loss component.
  
\subsection{Realistic situation and comparison with measurements}

\begin{figure}[!tb]
\centering
\includegraphics[scale=0.95]{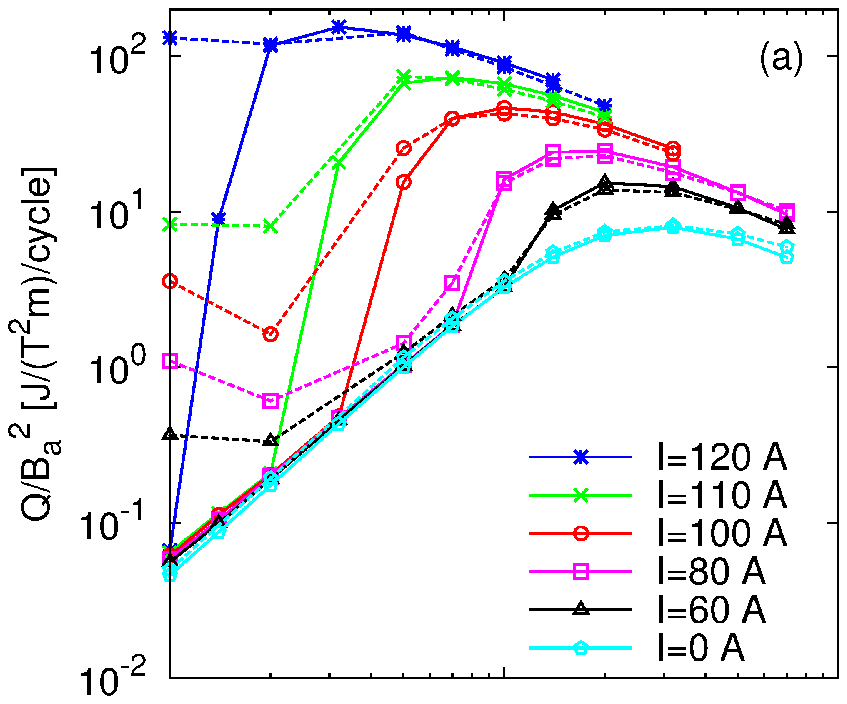}
\includegraphics[scale=0.95]{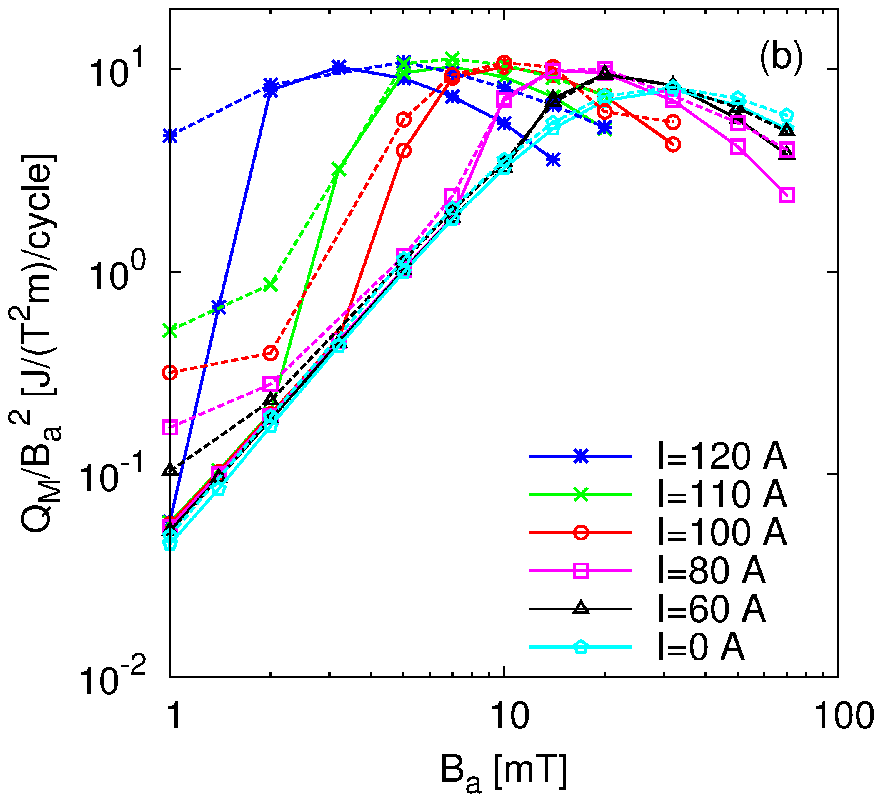}
\caption{(a) The total losses and (b) the magnetization losses obtained from the simulations for the experimental tape under several DC currents, $I$. The loss is normalized by the amplitude of the applied field squared. The solid and dashed lines represent losses obtained using MMEV and the $H$-formulation, respectively. The legend indicates the corresponding DC currents for both solid and dashed lines.}
\label{fig:LossesCSMECM}
\end{figure}

Having seen that the models do lead to different predictions in certain situations, we measured and simulated the coated conductor tape with measured ${\bf B}$-dependencies of $J_c$ and $n$ \cite{coatedIc} carrying a set of DC currents in a set of applied AC magnetic fields. The total AC losses and the magnetization losses obtained from the simulations are presented in figure~\ref{fig:LossesCSMECM}. Again, the DC current has been raised up to its amplitude value first in 5~ms, then it has been kept constant for 15~ms in zero applied field and then a 50~Hz sinusoidal ripple field perpendicular to the tape width has been applied on top of it. The losses have been integrated over the tenth cycle of the ripple field. In these cases, the hysteresis loops closed for all the cases to a reasonable extent, and we could separate the contribution of magnetization loss from the total loss also for the results obtained using ECM. For all the calculations, we have checked that there are no spurious effects due to the discrete meshing, since the results are not affected with increasing the number of elements. Moreover, the penetration depth of the critical region (for the CSM computations) expands over several elements from the edge, including the cases for the lowest applied AC fields. Similarly, for ECM computations, the current penetrates several elements from the edges of the tape.

\subsubsection{Comparison between the models}

Comparing the results, one observes that the qualitative and quantitative agreement between the predictions of the models are good in a wide range of situations (figure~\ref{fig:LossesCSMECM}). However, very large discrepancies occur again at the lowest AC fields of 1~mT amplitude, as was expected on the grounds of the previous section. We notice that ECM does not only predict higher magnetization losses but also especially higher transport losses than CSM in those cases.

\subsubsection{Measurements and comparison with the models}

\begin{figure}[!tb]
\centering
\includegraphics[scale=0.95]{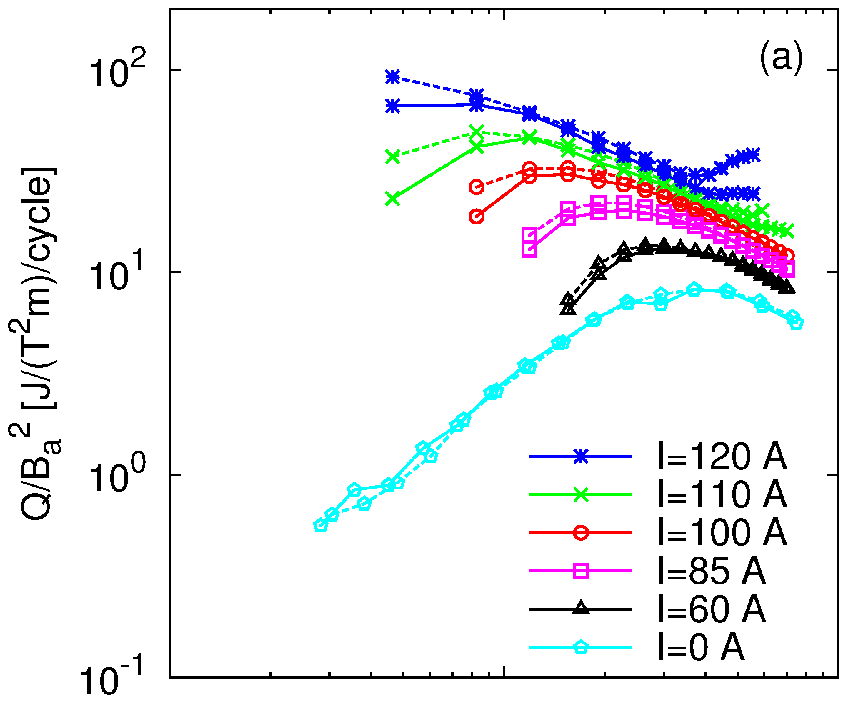}
\includegraphics[scale=0.95]{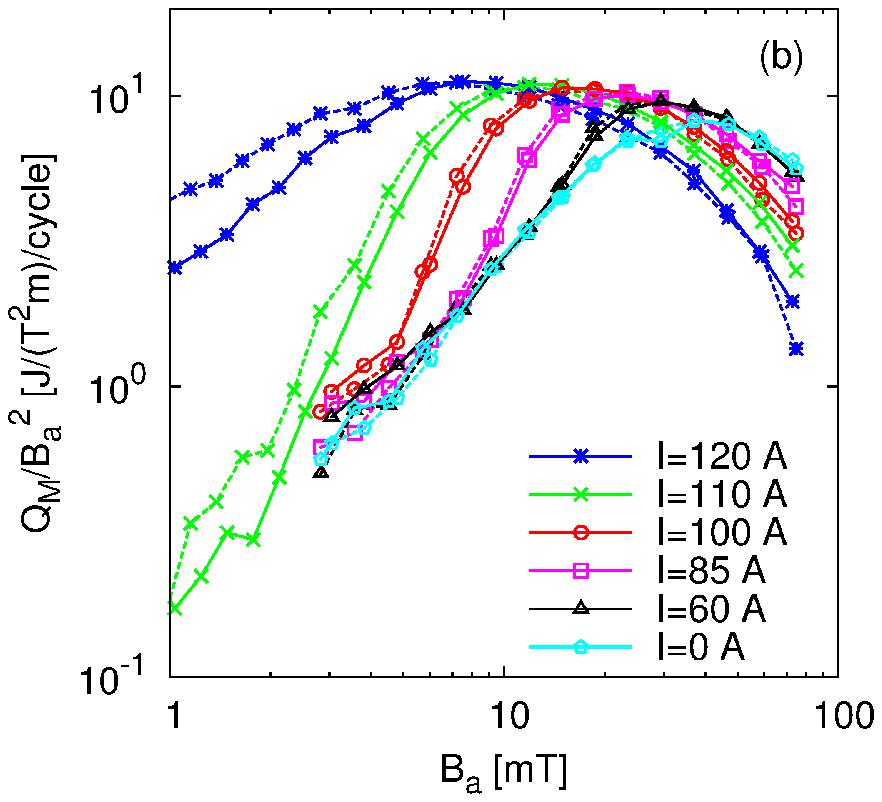}
\caption{The measured losses normalized with the applied field amplitude squared. The dashed and solid lines represent results at $f$~=~36~Hz and $f$~=~72~Hz, respectively. (a) The total losses interpolated and summed from measured magnetization loss and transport loss data. The total loss for $I = 0$~A has been taken to be merely the magnetization loss. (b) The measured magnetization losses.}
\label{fig:Measurements}
\end{figure}

The measured total losses and magnetization losses are presented in figure~\ref{fig:Measurements}. The total losses have been obtained by interpolating the measured magnetization and transport loss data and summing them up. Transport loss measurements for low currents and low fields were noisy, which is why we cannot say anything about the total losses in those cases, either. We performed the measurements for $f = 36$~Hz and $f = 72$~Hz, and a very slight frequency-dependence was observed, which we discuss two paragraphs below. The total loss increases with increasing the DC current for any AC applied field due to the dynamic magneto-resistance effect. In addition, the loss factor $Q/B_a^2$ as a function of the applied field amplitude $B_a$ presents a peak. The reasons are the following. First, the magnetization loss presents a peak at the saturation field. Second, the transport loss, which is due to the dynamic magneto-resistance effect, presents a certain onset. Above, the loss is proportional to $B_a$ (with $Q/B_a^2\propto 1/B_a$) and, below, the loss rapidly increases with $B_a$. For the highest DC currents and high $B_a$ the loss increases again with $B_a$. The cause is that the tape critical current decreases with the magnetic field. Then, at high $B_a$, the transport current is above the critical current at part or the whole AC cycle. As a consequence, the loss is mainly resistive with a loss per cycle that follows $Q=E_cI^{n+1}/(I_c^nf)$, from the power-law $E(J)$ relation of (\ref{power-law}). Consistently, the measured loss per cycle decreases with the frequency. Regarding the magnetization loss, at high AC applied field this loss contribution decreases with the transport current because it saturates part of the tape cross-section. However, the magnetization loss increases with the DC current at low applied fields, in agreement with \cite{ciszek02ACE}.

\begin{figure}[!tb]
\centering
\includegraphics[scale=0.95]{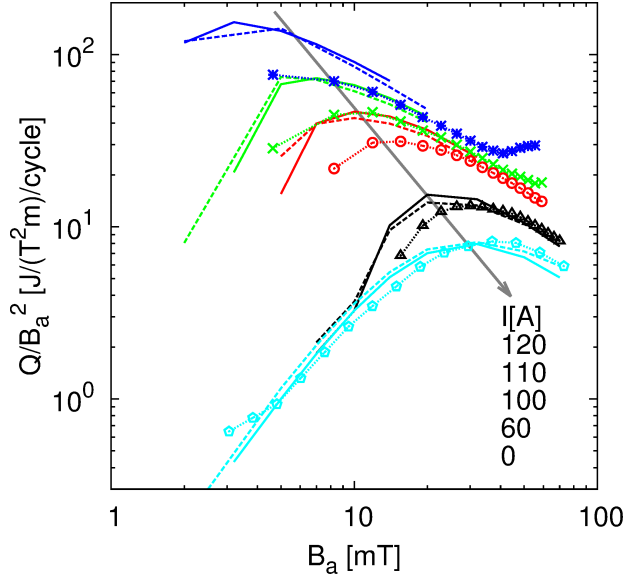}
\caption{Comparison of the measured total loss with simulations for $f$=50 Hz. Lines with symbols are for measurements, and solid and dashed lines are for simulations from MMEV and ECM, respectively. The DC current is $I$=120,110,100,60,0 A in the arrow direction. The measurements are linearly interpolated to 50 Hz from the acquired data at 36 and 72 Hz.}
\label{fig:Meastotcmp}
\end{figure}

When comparing the measured total losses with the simulation results of both models, one observes a satisfactory qualitative and quantitative agreement (figure \ref{fig:Meastotcmp}). However, the quantitative agreement is not perfect. For no DC current, both models agree well with the experiments, especially taking into account that the loss factor representation exaggerates the difference between curves compared to representing the loss directly. At very low applied fields, the measured curve presents a lower slope than the calculations. This is consistent with a possible degradation at the tape edges, although it could also be due to experimental error, since the measured signal was small at low $B_a$. The peak of the experimental contour is shifted to higher AC fields, although the curves for the measurements and the ECM calculations meet at higher AC field amplitudes. This suggests that the real $J_c$ is higher than that of the simulations for low magnetic fields. Notice that self-field correction on the $B$-dependence of $J_c$ is not straightforward, especially for magnetic fields below the tape self-field, around 20 mT \cite{coatedIc}. Under non-zero DC current, the agreement is very good at high $B_a$. With decreasing $B_a$, the difference between measurements and model predictions increases. This may also be attributed to an under-estimated $J_c$ at low magnetic fields. The agreement with experiments worsens with increasing the current because at high currents the AC loss is very sensitive to $I_c$, or the ratio $I/I_c$. Note that the models results for 100 and 110 A almost exaclty correspond to the measurements for 110 and 120 A, respectively. This is consistent with roughly 10 \% error of $J_c$ at low magnetic fields.

The frequency dependence may be explained as follows. By increasing the frequency, the average local $||E||$ increases, and so does $||J||$ due to the smooth $E$-$J$ relation. As a consequence, the effective critical current increases with the frequency and the dynamic magneto-resistance decreases. Since this is very sensitive to the critical current for $I$ close to $I_c$, there could be a significant frequency dependence for currents close to $I_c$, in consistence with the measurements. Another possible cause of frequency dependence is a certain current sharing with the stabilization layer. Since the DC voltage increases with the frequency, the portion of current in the stabilization layer also increases, causing loss decrease. This effect deserves further study and will be investigated in a future work.

%\begin{figure}[!tb]
%\centering
%\includegraphics[scale=0.95]{figure19a.ps}
%\includegraphics[scale=0.95]{figure19b.ps}
%\caption{Comparison of the measured magnetization loss with simulations for $f$=50 Hz. Lines with symbols are for measurements, and solid and dashed lines are for simulations from MMEV and ECM, respectively. Top: the DC current is 120,100,60 A in the arrow direction; bottom: DC current shown in the legend. The measurements are linearly interpolated to 50 Hz from the acquired data at 36 and 72 Hz.}
%\label{fig:Measmagcmp}
%\end{figure}

\begin{figure}[!tb]
\centering
\subfigure[]{
\includegraphics[scale=0.95]{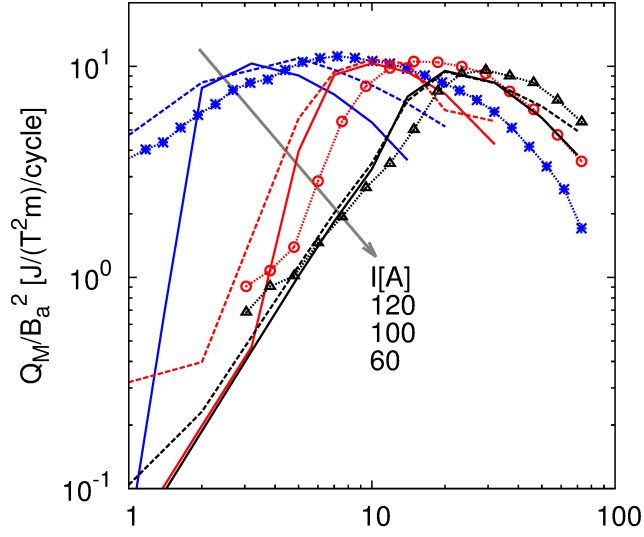}
\label{fig:Measmagcm1}
}
\subfigure[]{
\includegraphics[scale=0.95]{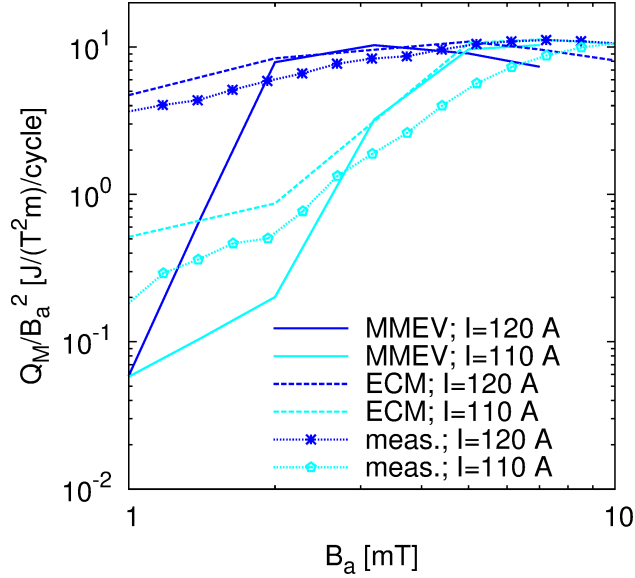}
\label{fig:Measmagcm2}
}
\caption{Comparison of the measured magnetization loss with simulations for $f$=50 Hz. Lines with symbols are for measurements, and solid and dashed lines are for simulations from MMEV and ECM, respectively. (a): the DC current is 120,100,60 A in the arrow direction; (b): DC current shown in the legend. The measurements are linearly interpolated to 50 Hz from the acquired data at 36 and 72 Hz.}
\label{fig:Measmagcm}
\end{figure}

In general, the agreement between the models and the measurements is better for the magnetization loss than for the total loss at the same DC current (figure \ref{fig:Measmagcmp}). The causes of discrepancy are the same as those for the total loss, except for the results at very low applied magnetic fields. Next, we discuss in detail the comparison with experiments for very low applied field amplitudes. To highlight the discrepancies of the results in low AC fields, the computed total and magnetization losses and measured magnetization losses for 1~mT AC field and 110 and 120~A DC currents have been listed in table~\ref{table:1mT_Losses}. Measurements for such low fields were relatively noisy, and the transport loss was not measured. However, the measured magnetization losses are of the same order of magnitude as the ones obtained using ECM. Especially for DC current of 120~A, which is very close to the measured self-field $I_\mathrm{c}$ of the tape (128~A), the CSM based computations have led to under-estimation of the magnetization loss, whereas ECM gives a better correspondence with the measurement. This can partially be explained by taking the $B$-dependence of the $n$-value into account in ECM. One should also note that there could be a degradation of $J_\mathrm{c}$ close to the tape edges, which could increase the measured loss at low amplitudes, but which was not taken into account in the simulations \cite{eucas10fmsc}. However, especially the total losses given by ECM are of completely different magnitude than the ones given by CSM. This, we believe, is again a consequence of the fact that the losses associated to DC components of the field and current are over-estimated in ECM, as discussed earlier. In particular, the transport losses are very high in ECM based results. We do not have the sufficient measurement data for transport losses, but we may calculate the threshold value of the magnetic field for the dynamic magneto-resistance effect to occur using~\eqref{Threshold}. For a case of a superconducting strip with a constant $I_{\mathrm{c}} = 120$~A and the same dimensions as the modelled tape, one obtains $B^* \approx 4$~mT for 100~A DC transport current, indicating that at AC fields of the order of 1~mT, practically no transport loss should occur, as CSM predicts. Of course, one cannot alone from this infer that the loss predictions of CSM are closer to reality in these cases, but more experiments are needed.

In summary, as we have seen, it is clear that the non-physical time-evolution of the current profile in ECM does affect the loss simulations, as well, which manifests that the power law resistivity does not describe the \emph{intrinsic} behaviour of a coated conductor tape. There could also be a degradation of $J_\mathrm{c}$ near the edges of the tape, which causes discrepancies between measurements and simulations, as it was not taken into account in the modelling. Still, CSM does not predict the sharp loss rise when increasing the DC current close to $I_\mathrm{c}$, whereas this is prominent in the ECM based results. CSM and ECM describe different features of the superconductor properties.

\begin{table}[!tb]
\begin{ruledtabular}
\caption{The computed total and magnetization losses and the measured magnetization losses for 1~mT AC fields and two different DC currents.}
\label{table:1mT_Losses}
\scriptsize 
\begin{tabular}{ p{1cm}  p{1.8cm}  p{1.8cm}  p{1.8cm}  p{1.8cm} p{1.8cm} }   
DC current [A] & Total loss (CSM) [J/m/cycle] & Total loss (ECM) [J/m/cycle] & Magnetization loss (CSM) [J/m/cycle] & Magnetization loss (ECM) [J/m/cycle] & Magnetization loss (measured 72 Hz) [J/m/cycle] \\
110 &		$6.393 \times 10^{-8}$		&	$8.290 \times 10^{-6}$	&		$5.691 \times 10^{-8}$		&	$5.151 \times 10^{-7}$		&	$1.85 \times 10^{-7}$			\\	
120 &		$7.157 \times 10^{-8}$		&	$1.314 \times 10^{-4}$	&   $5.747 \times 10^{-8}$  	&	$4.715 \times 10^{-6}$		&	$2.74 \times 10^{-6}$										\\   
\end{tabular}
\end{ruledtabular}
\end{table}

\section{Conclusions}

The appropriateness of CSM and ECM for modelling AC losses of DC biased superconductors in AC ripple fields has been investigated. The models have been compared with each other as well as measurements. For pure AC cases and DC-AC cases with the AC field significant enough compared to the DC field, the agreement between the models is good. In such situations, both models show good agreement qualitatively and quantitatively with measurements, as well. For the measurements, we have used a simple voltage tap configuration and demonstrated its correctness from electromagnetic theory.

For some of the studied cases, CSM and ECM yield different predictions for the behaviour of DC biased superconductors, reliable in different situations. These discrepancies are a consequence of the different $E$-$J$-relations used in the models: the finite $n$-value of ECM leads to losses also with a DC current or field. After setting a DC bias, a superconductor with power-law $E$-$J$ relation evolves towards a homogeneous current density profile by constantly redistributing the current density. This is not the behaviour observed in our Hall-probe mapping measurements, which present much slower relaxation effects and apparently reach a stationary state with CSM-like current distribution. These experiments suggest lower $E$ for the same $J$ than the power law $E$-$J$ relation, and therefore an $E$-$J$ relation closer to CSM for low $E$. However, further research on the topic is still required. Nonetheless, the results suggest that neither the power law used in ECM nor the sharp $E$-$J$-relation of CSM is the most appropriate local $E$-$J$-relation for high-temperature superconductors. 

In terms of AC loss simulations, ECM seems to over-estimate the losses related to low frequencies and DC. This is especially prominent for combinations of a very low AC field and a significantly large DC field compared to the AC field, as the stationary loss component of ECM becomes significant in such cases, leading to very long transients, too. Hence, there are also notable differences in the predictions of the models for instantaneous AC losses in terms of the shapes and magnitudes of $P(t)$ curves. Partially, they are related to the observation of over-estimated homogenization of current density, which is due to the stationary loss component, and partially they will require further investigation of the flux front patterns in the tapes. Furthermore, our measurement results suggest that the contribution of the magnetization losses to the total AC losses in DC biased superconductors subject to ripple fields is closer to the prediction given by ECM.

Nonetheless, more high-resolution experiments are needed for more information about the real behaviour of high-temperature superconductors under such conditions. In particular, studying a single tape under simultaneous DC applied magnetic field and DC current is suggested. Further work should also study the effect of extending the relaxation time (up to minutes or hours) on the AC loss predictions using the power law $E$-$J$ relation, as the transients can be extremely long: the steady-state was not reached for the cases of lowest applied fields in our simulations, because of the slowly descending stationary loss component in ECM.

In addition to the observations directly concerning AC loss simulations, the results also suggest a deeper insight about modelling: while it would be tempting to claim the contrary, neither CSM or ECM are able to fully reflect the intrinsic properties of superconductors. Further research on this topic is not only important for predicting AC losses in applications, but also for the fundamental quest for the most appropriate local $E$-$J$-relation for high-temperature superconductors.

% If in two-column mode, this environment will change to single-column format so that long equations can be displayed. 
%\begin{widetext}
%$$\mbox{put long equation here}$$
%\end{widetext}

% Figures should be put into the text as floats. 
% Use the graphics or graphicx packages (distributed with LaTeX2e).
% Put the label that you will use with \ref{} command in the braces of the \label{} command.
%
% \begin{figure}
% \end{figure}

% \begin{table}
% \end{tabular}
% \end{table}

\begin{acknowledgments}

This work was partially supported by The Academy of Finland project $\#$250652. We also acknowledge the support of EURATOM FU-CT-2007-00051 project co-funded by the Slovak Research and Development Agency under contract number DO7RP-0018-12.

\end{acknowledgments}

%\section*{References}

%\bibliographystyle{unsrt}	
%\bibliography{all}

% JAP Requires to write the references by hand, or copy-paste the bbl file here. It does not accept .bib files.

% Footnotes go to bibliography as well

\end{document}